\documentclass[12pt,english]{article}
\usepackage{tgtermes}

\usepackage[T1]{fontenc}
\usepackage[latin9]{inputenc}
\usepackage[a4paper]{geometry}
\usepackage{verbatim}
\usepackage{rotfloat}
\usepackage{amsmath}
\usepackage{amssymb}
\usepackage{graphicx}
\usepackage[authoryear]{natbib}

\makeatletter

\providecommand{\tabularnewline}{\\}

\usepackage{babel}

\makeatother

\usepackage{babel}
\begin{document}

\title{Bayesian Singular Value Regularization \linebreak{}
 via a Cumulative Shrinkage Process}

\author{Masahiro Tanaka\thanks{Department of Economics, Kanto Gakuen University; Graduate School
of Economics, Waseda University. Address: 200, Fujiagucho, Ohta, Gunma
169-8050 Japan. Email: gspddlnit45@toki.waseda.jp. Personal website:
https://sites.google.com/view/masahirotanakastat}}

\date{October 8, 2020}

\maketitle
 
\begin{abstract}
This study proposes a novel hierarchical prior for inferring possibly
low-rank matrices measured with noise. We consider three-component
matrix factorization, as in singular value decomposition, and its
fully Bayesian inference. The proposed prior is specified by a scale
mixture of exponential distributions that has spike and slab components.
The weights for the spike/slab parts are inferred using a special
prior based on a cumulative shrinkage process. The proposed prior
is designed to increasingly aggressively push less important, or essentially
redundant, singular values toward zero, leading to more accurate estimates
of low-rank matrices. To ensure the parameter identification, we simulate
posterior draws from an approximated posterior, in which the constraints
are slightly relaxed, using a No-U-Turn sampler. By means of a set
of simulation studies, we show that our proposal is competitive with
alternative prior specifications and that it does not incur significant
additional computational burden. We apply the proposed approach to
sectoral industrial production in the United States to analyze the
structural change during the Great Moderation period.

\bigskip{}
 
\end{abstract}


\section{Introduction}

The problem of low-rank matrix estimation has been receiving increasing
attention in natural and social sciences and other fields. In many
applications, the data approximately exist in a low-dimensional linear
subspace. Low-rank matrix factorization (decomposition) is often used
to recover the low-rank structure, remove measurement noise, and complete
missing entries (see, e.g., \citealp{Candes2010,Shi2017} for a survey
of the literature). Low-rank matrix factorization serves several purposes,
such as dimension reduction, data imputation, and structural analysis.
It also has a close ties to more elaborate models for high-dimensional
data, e.g., static factor models \citep{Geweke1996}, reduced rank
regression \citep{Geweke1996a}, and factor regression \citep{West2003}.
As discussed in \citet{Tanaka2020}, matrix completion can also be
used for causal inference.

Assessment of the uncertainty in inference is an integral component
of scientific research and many practical applications. Non-Bayesian
approaches to matrix factorization struggle to quantify uncertainty
(e.g., \citealp{Chen2019}). In contrast, fully Bayesian approaches
can straightforwardly obtain credible sets as a byproduct of posterior
simulation. Therefore, there is a good reason to employ a fully Bayesian
approach in matrix factorization.

In this study, we consider fully Bayesian inference of three-component
factorization of matrices, as in singular value decomposition. A noisy
matrix $\boldsymbol{Y}$ is modeled as 
\[
\boldsymbol{Y}=\boldsymbol{\Phi}\boldsymbol{\Omega}\boldsymbol{\Psi}^{\top}+\boldsymbol{U},
\]
where $\boldsymbol{U}$ is a matrix of measurement errors. Moreover,
some elements of $\boldsymbol{Y}$ may be missing. Two types of constraints
are needed for the exact identification of the matrix factorization.
First, $\boldsymbol{\Phi}$ and $\boldsymbol{\Psi}$ are supposed
to be unitary (or orthonormal) matrices, i.e., $\boldsymbol{\Phi}^{\top}\boldsymbol{\Phi}=\boldsymbol{\Psi}^{\top}\boldsymbol{\Psi}=\boldsymbol{I}$.
Second, $\boldsymbol{\Omega}$ is assumed to be diagonal, and its
diagonal elements are nonnegative and arranged in order of size.

In the non-Bayesian literature, low-rank matrix estimation/completion
using the nuclear norm penalty has been intensively studied (e.g.,
\citealp{Keshavan2010,Rohde2011,Negahban2011,Koltchinskii2011,Klopp2014,Gaiffas2017}).
This strategy amounts to estimation of a matrix while penalizing based
on its rank since the nuclear norm of a matrix is a convex relaxation
of its rank \citep{Fazel2001}. An estimator using the nuclear norm
penalty coincides with the posterior mode under two prior specifications.
First, for three-component factorization, an exponential prior for
the singular values, the diagonal elements of $\boldsymbol{\Omega}$,
corresponds to the nuclear norm penalty. Second, for two-component
factorization, the use of an independent normal prior is tantamount
to the use of the nuclear norm penalty (e.g., \citealp{Salakhutdinov2008,Babacan2012,Fazayeli2014}).

The main contribution of this study is to propose a novel shrinkage
prior for the diagonal elements of $\boldsymbol{\Omega}$, i.e., the
singular values. The proposed prior is specified by a scale mixture
of exponential distributions with spike and slab components. The spike
part is given by an exponential distribution with a large rate parameter;
that is, both the mean and variance are close to zero. The slab part
is specified as a scale mixture of exponential distributions whose
mixing distribution is a gamma distribution; thus, the slab part is
distributed according to a Lomax distribution. The distribution of
the weights of the spike part is specified based on a cumulative shrinkage
process \citet{Legramantiforthcoming}. This prior specification of
weights is designed to increasingly shrink less important, or essentially
redundant, singular values toward zero.

By means of a set of Monte Carlo experiments, we demonstrate that
in terms of estimation accuracy, the proposed prior is competitive
with alternative choices under various environments. Our proposal
is especially beneficial in difficult situations, for example, when
the true rank of a matrix is small relative to its size and many entries
are missing. While the shrinkage priors including the proposed prior
have biases in major singular values, this cost is overwhelmed by
the gains from pushing minor singular values toward zero. In regard
to this point, the CSPE prior is the best among the alternatives we
consider, as it aggressively shrinks insignificant singular values.

To demonstrate the proposed prior, we apply it to sectoral industrial
production in the United States. The purpose of this application is
to analyze the Great Moderation period, which represented a decrease
in the volatility of major macroeconomic indicators after the mid-1980s
in the United States and other developed countries (e.g., \citealp{McConnell2000,Summers2005}).
Our application is intended to shed new light on this issue by inferring
the singular values of a panel of sectoral industrial production growth
for different periods. We find that compared to that in the period
before the Great Moderation, in the period during the Great Moderation,
the relative contributions of the large singular values to the observed
variability are significantly smaller, while the relative contributions
of the small singular values and the measurement errors, or idiosyncratic
shocks, are more pronounced.

The remainder of this paper is structured as follows. Section 2 introduces
the proposed approach. Section 3 conducts a set of simulation studies
to compare the proposed approach with other alternatives. Section
4 applies the proposed approach to real data. Finally, Section 5 concludes
the paper.

\section{Model and Inference}

\subsection{Model}

We infer a $J$-by-$T$ possibly rank-deficient matrix observed with
additive noise: 
\[
\boldsymbol{Y}=\boldsymbol{\Theta}+\boldsymbol{U},
\]
where $\boldsymbol{Y}=\left(y_{j,t}\right)$ is a $J$-by-$T$ matrix
of observations, $\boldsymbol{\Theta}=\left(\theta_{j,t}\right)$
is a $J$-by-$T$ matrix to be inferred, and $\boldsymbol{U}=\left(u_{j,t}\right)$
is a $J$-by-$T$ matrix of normally distributed measurement errors
with precision $\tau$, $u_{j,t}\sim\mathcal{N}\left(0,\tau^{-1}\right)$.\footnote{$\boldsymbol{A}=\left(a_{j,t}\right)$ denotes a matrix whose generic
$\left(j,t\right)$-element is $a_{j,t}$.} To induce $\boldsymbol{\Theta}$ to be (approximately) rank deficient,
we decompose $\boldsymbol{\Theta}$ into three parts

\[
\boldsymbol{\Theta}=\boldsymbol{\Phi}\boldsymbol{\Omega}\boldsymbol{\Psi}^{\top},\quad\boldsymbol{\Omega}=\textrm{diag}\left(\omega_{1},...,\omega_{K}\right),
\]
where the sizes of $\boldsymbol{\Phi}$, $\boldsymbol{\Omega}$, and
$\boldsymbol{\Psi}$ are $J$-by-$K$, $K$-by-$K$, and $T$-by-$K$,
respectively. When $\boldsymbol{Y}$ is incompletely observed, missing
elements are completed via data augmentation \citep{Tanner1987},
more generally, Gibbs sampling. Furthermore, $\boldsymbol{\Theta}$
and missing entries in $\boldsymbol{Y}$ are alternately simulated
from the conditional posteriors.

Unlike previous studies, we ensure the identification of the unknown
parameters, $\boldsymbol{\Phi}$, $\boldsymbol{\Omega}$, and $\boldsymbol{\Psi}$.
Indeed, the exact identification is not necessary for posterior analysis
of $\boldsymbol{\Theta}$, but in general, when unknown parameters
are not identified, the posterior simulation is numerically inefficient
or can even diverge in some situations.\footnote{\citet{Tang2019} and \citet{Tanaka2020} use posterior simulators
that account for this problem, but they do not resolve it entirely. }.

For parameter identification, we impose three types of constraints.
First, we choose $K$ such that the number of unknown parameters in
the factorization $\boldsymbol{\Phi}\boldsymbol{\Omega}\boldsymbol{\Psi}^{\top}$
does not exceed that in the original matrix $\boldsymbol{\Theta}$:\footnote{See \citet{Chan2018} for a related discussion.}
\[
JK+K+TK\leq JT,
\]
or equivalently, 
\[
K\leq\frac{JT}{J+1+T}.
\]
Second, $\boldsymbol{\Phi}$ and $\boldsymbol{\Psi}$ are supposed
to be unitary over $K$, $\boldsymbol{\Phi}^{\top}\boldsymbol{\Phi}=\boldsymbol{\Psi}^{\top}\boldsymbol{\Psi}=\boldsymbol{I}_{K}$.
While these constraints keep the magnitudes of $\boldsymbol{\Phi}$
and $\boldsymbol{\Psi}$ constant, the column signs of $\boldsymbol{\Phi}$
and $\boldsymbol{\Psi}$ are not identified. We address this issue
by modifying a posterior simulator, as discussed subsequently. Third,
the diagonal elements of $\boldsymbol{\Omega}$ are assumed to be
nonnegative and arranged in descending order, $\omega_{1}\geq\omega_{2}\geq\cdots\geq\omega_{K}\geq0$,
which can also be written as 
\begin{eqnarray*}
\omega_{1}-\omega_{2} & \geq & 0,\\
\omega_{2}-\omega_{3} & \geq & 0,\\
 & \vdots\\
\omega_{K-1}-\omega_{K} & \geq & 0,\\
\omega_{K} & \geq & 0.
\end{eqnarray*}
Let $\boldsymbol{\omega}=\left(\omega_{1},...,\omega_{K}\right)^{\top}$.
Then, we have the following matrix representation: 
\begin{equation}
\boldsymbol{C}\boldsymbol{\omega}\geq\boldsymbol{0}_{K},\quad\textrm{with}\;\boldsymbol{C}=\left(\begin{array}{ccccc}
1 & -1 & 0 & \cdots & 0\\
0 & 1 & -1 & \ddots & \vdots\\
0 & 0 & \ddots & \ddots & 0\\
\vdots & \ddots & \ddots & 1 & -1\\
0 & \cdots & 0 & 0 & 1
\end{array}\right).\label{eq: inequality constraints}
\end{equation}
Let $\mathcal{C}$ denote a real space that satisfies the constraints
$\boldsymbol{\omega}$ provided by (\ref{eq: inequality constraints}),
$\boldsymbol{\omega}\in\mathcal{C}$.

\subsection{Priors}

The prior distribution of $\omega_{k}$ is specified by a scale mixture
of exponential distributions that has spike and slab components. Following
\citet{Legramantiforthcoming}, the weights for the spike part $\pi_{k}$
follow the stick-breaking construction of the Dirichlet process (e.g.,
\citealp{Ishwaran2001}). Ignoring the constraints (\ref{eq: inequality constraints}),
the prior of $\boldsymbol{\omega}$ is given by the following hierarchy:
\begin{eqnarray}
\omega_{k}|\lambda_{k} & \sim & \mathcal{E}\left(\lambda_{k}\right),\quad k=1,...,K,\nonumber \\
\lambda_{k}|\pi_{k} & \sim & \left(1-\pi_{k}\right)\mathcal{G}\left(\kappa_{1},\kappa_{2}\right)+\pi_{k}\delta,\quad k=1,...,K,\label{eq: prior of lambda}\\
\pi_{k} & = & \sum_{l=1}^{k}\gamma_{l},\quad\text{with }\gamma_{l}=\upsilon_{l}\prod_{m=1}^{l-1}\left(1-\upsilon_{m}\right),\quad k=1,...,K,\nonumber \\
\upsilon_{m} & \sim & \mathcal{B}\left(1,\alpha\right),\quad m=1,...,K-1,\quad\upsilon_{K}=1,\nonumber 
\end{eqnarray}
where $\delta$, $\kappa_{1}$ and $\kappa_{2}$ are prefixed hyperparameters,
$\mathcal{E}\left(a\right)$ denotes an exponential distribution with
rate parameter $a$, $\mathcal{G}\left(a,b\right)$ denotes a gamma
distribution with shape parameter $a$ and rate parameter $b$, and
$\mathcal{B}\left(a,b\right)$ denotes a beta distribution (of the
first kind) with mean $a/\left(a+b\right)$. In the following, we
call this hierarchical prior a cumulative shrinkage process exponential
(CSPE) prior. The weights $\pi_{1},...,\pi_{K}$ are arranged in ascending
order, $\pi_{1}\leq\pi_{2}\leq\cdots\leq\pi_{K}$, while $\omega_{1},...,\omega_{K}$
are arranged in descending order, $\omega_{1}\geq\omega_{2}\geq\cdots\geq\omega_{K}$.
Therefore, relatively less important singular values are more likely
to belong to the spike part.

The marginal prior of $\omega_{k}$ for the slab part is represented
by a Lomax distribution; integrating out $\lambda_{k}$ yields 
\[
\omega_{k}|\pi_{k}\sim\left(1-\pi_{k}\right)\mathcal{L}\left(\kappa_{1},\kappa_{2}\right)+\pi_{k}\mathcal{E}\left(\delta\right),\quad k=1,...,K,
\]
where $\mathcal{L}\left(a,b\right)$ is a Lomax distribution with
shape $a$, scale $b$, and probability density function (PDF) 
\[
p\left(x\right)=\frac{a}{b}\left(1+\frac{x}{b}\right)^{-\left(a+1\right)}.
\]
The Lomax distribution is a type I Pareto distribution that has nonnegative
support; thus, it has a heavier tail than the exponential distribution.
The mean of the slab part is $\kappa_{2}/\left(\kappa_{1}-1\right)$
for $\kappa_{1}>1$ and is undefined otherwise, and the variance is
$\kappa_{2}^{2}\kappa_{1}\left(\kappa_{1}-1\right)^{-2}\left(\kappa_{1}-2\right)^{-1}$
for $\kappa_{1}>2$, $\infty$ for $1<\kappa_{1}\leq2$, and undefined
otherwise. %
\begin{comment}
When we choose $\kappa_{1}=2+10^{-b}$ with $b>0$, the mean and variance
of the slab part are approximately $\kappa_{2}$ and $2\kappa_{2}\times10^{b}$,
respectively. 
\end{comment}

The rate parameter for the spike part $\delta$ is set to a sufficiently
large value such that both the mean $\left(\delta^{-1}\right)$ and
variance $\left(\delta^{-2}\right)$ of the spike part are small,
thereby pushing nonsignificant singular values toward zero. Although
$\delta$ can be infinite, we can improve the computational efficiency
of the posterior simulation by setting it to a finite value \citep{Ishwaran2005}.
Our default choice is $\delta=10$, which implies that the mean and
variance of the spike part are 0.1 and 0.01, respectively.

The hyperparameters for the slab part, $\kappa_{1}$ and $\kappa_{2}$,
are chosen to make this component sufficiently noninformative. Elicitation
of $\kappa_{1}$ and $\kappa_{2}$ is not trivial because, for most
applications, we have no prior knowledge about $\boldsymbol{\omega}$.
In this paper, our default choice is $\kappa_{1}=2$ and $\kappa_{2}=20$,
which implies that the mean and variance of the slab part are 20 and
$\infty$, respectively. The PDFs of the spike and slab parts under
the default choices are shown in Figure 1. By setting $\kappa_{2}$
to a large value, we can make the slab part arbitrarily noninformative,
but at least for the simulation environments in the subsequent section,
we find no significant gain from doing so. Thus, we recommend setting
$\kappa_{2}$ to a moderate value, thereby allowing the slab part
to play a role in shrinkage.

Tuning $\alpha$ is also nontrivial. When some prior belief about
the rank of $\boldsymbol{\Theta}$ is available, $\alpha$ can be
chosen systematically. The expectation of $\pi_{k}$ is 
\[
\mathbb{E}\left[\pi_{k}\right]=1-\frac{\alpha^{k}}{\left(1+\alpha\right)^{k}},\quad k=1,...,K.
\]
We can tune $\alpha$ such that the probability of the $k$th singular
value being insignificant (i.e., belonging to the spike part) is set
to a target value $q\in\left(0,1\right)$, 
\[
q=1-\frac{\alpha^{k}}{\left(1+\alpha\right)^{k}}.
\]
By solving this equation, we obtain 
\[
\alpha=\frac{\left(1-q\right)^{\frac{1}{k}}}{1-\left(1-q\right)^{\frac{1}{k}}}.
\]
Figure 2 depicts examples of prior means of $\pi_{k}$ and $1-\pi_{k}$
for $K=14$. Panel (a) is an example of a conservative choice, $\alpha\approx18.25$,
where the prior mean of the $\left(K-1\right)$th singular value being
insignificant is 0.5. In turn, Panel (b) is an example of an aggressive
choice, $\alpha\approx5.16$, where the prior mean of the $\left(K-1\right)$th
singular value being insignificant is 0.9.

For $\boldsymbol{\Phi}$ and $\boldsymbol{\Psi}$, we use uniform
Haar priors. The prior densities of $\boldsymbol{\Phi}$ and $\boldsymbol{\Psi}$
are represented as 
\[
p\left(\boldsymbol{\Phi}\right)\propto\mathbb{I}\left(\boldsymbol{\Phi}\in\mathcal{M}_{J\times K}\right),\quad p\left(\boldsymbol{\Psi}\right)\propto\mathbb{I}\left(\boldsymbol{\Psi}\in\mathcal{M}_{T\times K}\right),
\]
where $\mathcal{M}_{A\times B}=\left\{ \boldsymbol{X}\in\mathbb{R}^{A\times B}:\;\boldsymbol{X}^{\top}\boldsymbol{X}=\boldsymbol{I}_{B}\right\} $
denotes the Stiefel manifold with dimensions of $A$-by-$B$ and $\mathbb{I}\left(\cdot\right)$
denotes the indicator function. The error precision parameter $\tau$
is inferred using a gamma prior, $\tau\sim\mathcal{G}\left(\nu_{1},\nu_{2}\right)$,
where $\nu_{1}$ and $\nu_{2}$ are shape and rate parameters, respectively.

\subsection{Posterior simulation}

We conduct a posterior simulation using a hybrid of two algorithms.
We sample $\boldsymbol{\Phi}$, $\boldsymbol{\Psi}$, and $\boldsymbol{\omega}$
using a No-U-Turn sampler \citep{Hoffman2014}, which is an adaptive
version of the Hamiltonian Monte Carlo method \citep{Duane1987,Neal2011}.
The remaining parameters are simulated using a Gibbs sampler.

To address the inequality constraints (\ref{eq: inequality constraints}),
we transform $\boldsymbol{\omega}$ as $\boldsymbol{\omega}^{*}=\boldsymbol{C}\boldsymbol{\omega}$
and sample $\boldsymbol{\omega}^{*}\geq\boldsymbol{0}_{K}$ from the
$K$-dimensional nonnegative real space $\mathbb{R}_{+}^{K}$. This
transformation does not require any adjustment of the posterior estimates
since the Jacobian determinant of $\boldsymbol{\omega}^{*}$ is always
equal to one, 
\[
\left|\left(\frac{\partial\boldsymbol{\omega}^{*}}{\partial\boldsymbol{\omega}}\right)^{-1}\right|=\left|\boldsymbol{C}^{-1}\right|=1.
\]

Given the transformation, the indicator function that represents the
constraints on $\boldsymbol{\omega}$, $\mathbb{I}\left(\boldsymbol{\omega}\in\mathcal{C}\right)$,
is replaced by the indicator function that represents the inequality
constraints on $\boldsymbol{\omega}^{*}$, $\mathbb{I}\left(\boldsymbol{\omega}^{*}\in\mathbb{R}_{+}^{K}\right)$,
while the the (unconstrained) prior density of $\boldsymbol{\omega}$,
$p\left(\boldsymbol{\omega}\right)$, is evaluated as a function of
$\boldsymbol{\omega}^{*}$, $p\left(\boldsymbol{\omega}\left(\boldsymbol{\omega}^{*}\right)\right)$,
where $\boldsymbol{\omega}\left(\boldsymbol{\omega}^{*}\right)=\boldsymbol{C}^{-1}\boldsymbol{\omega}^{*}$
denotes a backward transformation.

To simulate $\boldsymbol{\vartheta}=\left(\textrm{vec}\left(\boldsymbol{\Phi}\right)^{\top},\textrm{vec}\left(\boldsymbol{\Psi}\right)^{\top},\left(\boldsymbol{\omega}^{*}\right)^{\top}\right)^{\top}$
efficiently, the constraints are relaxed slightly \citep{Duan2020}.
For $\boldsymbol{\Phi}$ and $\boldsymbol{\Psi}$, we define the relaxation
functions as

\begin{eqnarray*}
\mathbb{I}\left(\boldsymbol{\Phi}\in\mathcal{M}_{J\times K}\right)\approx\iota_{\eta_{1}}\left(\boldsymbol{\Phi}\right) & = & \exp\left[-\eta_{1}\left\Vert \boldsymbol{\Phi}^{\top}\boldsymbol{\Phi}-\boldsymbol{I}_{K}\right\Vert _{F}^{2}\right],
\end{eqnarray*}

\[
\mathbb{I}\left(\boldsymbol{\Psi}\in\mathcal{M}_{T\times K}\right)\approx\iota_{\eta_{1}}\left(\boldsymbol{\Psi}\right)=\exp\left[-\eta_{1}\left\Vert \boldsymbol{\Psi}^{\top}\boldsymbol{\Psi}-\boldsymbol{I}_{K}\right\Vert _{F}^{2}\right],
\]
where $\eta_{1}\left(\geq0\right)$ is a hyperparameter and $\left\Vert \cdot\right\Vert _{F}$
denotes the Frobenius norm. The inequality constraints on $\boldsymbol{\omega}^{*}$
are approximated using a sigmoid function:

\[
\mathbb{I}\left(\boldsymbol{\omega}^{*}\in\mathbb{R}_{+}^{K}\right)\approx\rho_{\eta_{2}}\left(\boldsymbol{\omega}^{*}\right)=\prod_{k=1}^{K}\frac{1}{1+\exp\left(-\eta_{2}\omega_{k}^{*}\right)},
\]
where $\eta_{2}\left(\geq0\right)$ is a hyperparameter. Thus, the
conditional posterior of $\boldsymbol{\vartheta}$ is approximated
as 
\begin{eqnarray*}
p\left(\boldsymbol{\vartheta}|\boldsymbol{Y};\tau,\boldsymbol{\lambda}\right) & \propto & p\left(\boldsymbol{Y}|\boldsymbol{\Phi},\boldsymbol{\Psi},\boldsymbol{\omega}\left(\boldsymbol{\omega}^{*}\right),\tau\right)\mathbb{I}\left(\boldsymbol{\Phi}\in\mathcal{M}_{J\times K}\right)\mathbb{I}\left(\boldsymbol{\Psi}\in\mathcal{M}_{T\times K}\right)\mathbb{I}\left(\boldsymbol{\omega}^{*}\in\mathbb{R}_{+}^{K}\right)p\left(\boldsymbol{\omega}\left(\boldsymbol{\omega}^{*}\right)|\boldsymbol{\lambda}\right)\\
 & \approx & p\left(\boldsymbol{Y}|\boldsymbol{\Phi},\boldsymbol{\Psi},\boldsymbol{\omega}\left(\boldsymbol{\omega}^{*}\right),\tau\right)\iota_{\eta_{1}}\left(\boldsymbol{\Phi}\right)\iota_{\eta_{1}}\left(\boldsymbol{\Psi}\right)\rho_{\eta_{2}}\left(\boldsymbol{\omega}^{*}\right)p\left(\boldsymbol{\omega}\left(\boldsymbol{\omega}^{*}\right)|\boldsymbol{\lambda}\right).
\end{eqnarray*}
As the column signs of $\boldsymbol{\Phi}$ and $\boldsymbol{\Psi}$
are not identified, in each iteration, we flip the signs of the columns
of $\boldsymbol{\Phi}$ and $\boldsymbol{\Psi}$ if the signs of the
first elements of the columns of $\boldsymbol{\Phi}$ do not coincide
with those of the initial values. The gradient of the log approximate
conditional posterior of $\boldsymbol{\vartheta}$ can then be analytically
evaluated (see the Appendix).

While $\eta_{1}$ and $\eta_{2}$ are supposed to be large values,
we can improve the computational efficiency by relaxing the constraints
at the beginning of the posterior simulation and then tightening them
gradually to a prespecified level. For the $i$th MCMC iteration,
$\eta_{l}$ is set as 
\[
\eta_{l}=\begin{cases}
\bar{\eta}_{l}\left[1-\exp\left(\frac{\log\left(1-a_{\eta}\right)}{b_{\eta}\bar{\eta}_{l}}\times i\right)\right], & i<\bar{\eta}_{l}\\
\bar{\eta}_{l}, & i\geq\bar{\eta}_{l}
\end{cases},\quad l=1,2,
\]
where $\bar{\eta}_{l}$ denotes a prespecified upper bound and $a_{\eta}$
and $b_{\eta}$ are tuning parameters. This function amounts to the
product of $\bar{\eta}_{l}$ and the cumulative probability function
of an exponential distribution where the rate parameter is chosen
such that $\eta_{l}$ reaches $a_{\eta}\bar{\eta}_{l}$ in the $\left(b_{\eta}\bar{\eta}_{l}\right)$th
iteration. In this paper, we choose $\bar{\eta}_{1}=\bar{\eta}_{2}=10^{3}$,
$a_{\eta}=0.5$, and $b_{\eta}=0.1$.

The remaining parameters are updated via Gibbs steps. We introduce
the auxiliary indicator variable $z_{k}$ and rewrite the conditional
priors of $\lambda_{k}$ \eqref{eq: prior of lambda} as
\[
\lambda_{k}|\pi_{k}\sim\left(1-\mathbb{I}\left(z_{k}\leq k\right)\right)\mathcal{G}\left(\kappa_{1},\kappa_{2}\right)+\mathbb{I}\left(z_{k}\leq k\right)\delta,\quad k=1,...,K.
\]
Then, $z_{k}$ can be sampled from the following conditional posterior
distributions: 
\[
p\left(z_{k}=l|\text{rest}\right)\propto\begin{cases}
\pi_{l}f_{E}\left(\omega_{k}|\delta\right), & l=1,...,k,\\
\pi_{l}f_{L}\left(\omega_{k}|\kappa_{1},\kappa_{2}\right), & l=k+1,...,K,
\end{cases}
\]
where ``rest'' denotes random variables other than $z_{k}$ and
$\boldsymbol{Y}$,\footnote{For instance, the ``rest'' for $z_{k}$ is a set composed of $\boldsymbol{\Phi}$,
$\boldsymbol{\Psi}$, $\boldsymbol{\omega}$, $\upsilon_{1},...,\upsilon_{K}$,
$\lambda_{1},...,\lambda_{K}$, $\tau$, and $\boldsymbol{Y}$.} $f_{E}\left(x|a\right)$ denotes the PDF of an exponential distribution
with rate $a$ evaluated at $x$, and $f_{L}\left(x|a,b\right)$ denotes
the PDF of a Lomax distribution with shape $a$ and rate $b$. $\upsilon_{k}$
and $\lambda_{k}$ are simulated from the conditional posteriors,
which are specified respectively as

\begin{eqnarray*}
\upsilon_{k}|\text{rest} & \sim & \mathcal{B}\left(1+\sum_{l=1}^{K}\mathbb{I}\left(z_{l}=k\right),\;\alpha+\sum_{l=1}^{K}\mathbb{I}\left(z_{l}>k\right)\right),\quad k=1,...,K-1,\\
\lambda_{k}|\text{rest} & \sim & \mathbb{I}\left(z_{k}\leq k\right)\delta+\left(1-\mathbb{I}\left(z_{k}\leq k\right)\right)\mathcal{G}\left(\kappa_{1}+1,\;\kappa_{2}+\omega_{k}\right),\quad k=1,...,K.
\end{eqnarray*}
The conditional posteriors of $\tau$ and the missing elements of
$\boldsymbol{Y}$ (if any) are 
\begin{eqnarray*}
\tau|\text{rest} & \sim & \mathcal{G}\left(\nu_{1}+\frac{JT}{2},\;\nu_{2}+\frac{1}{2}\textrm{vec}\left(\boldsymbol{Y}-\boldsymbol{\Theta}\right)^{\top}\textrm{vec}\left(\boldsymbol{Y}-\boldsymbol{\Theta}\right)\right),\\
y_{j,t}|\text{rest} & \sim & \mathcal{N}\left(\theta_{j,t},\tau^{-1}\right),\quad\left(j,t\right)\in\mathcal{I}_{miss},
\end{eqnarray*}
where $\mathcal{I}_{miss}$ is a set of indices $\left(j,t\right)$
that corresponds to the missing elements. The Appendix summarizes
a single cycle of the posterior simulator.

\subsection{Relation to the existing literature}

The exponential prior is a special case of the proposed prior with
$\lambda=\lambda_{k}$, $\forall k$. The exponential prior has close
ties to the nuclear norm penalty, which has been extensively studied
in the non-Bayesian literature on matrix estimation/completion (e.g.,
\citealp{Keshavan2010,Rohde2011,Negahban2011,Koltchinskii2011,Klopp2014,Gaiffas2017}).
As shown in \citet{Fazel2001}, the nuclear norm penalty is a convex
relaxation of the rank of a matrix; thus, the nuclear norm penalty
acts as a penalty on the rank of a matrix. Furthermore, as discussed
in the abovementioned studies, the nuclear norm penalty has favorable
theoretical properties, and the choice of the tuning parameter, which
corresponds to the rate parameter of the exponential prior, is not
trivial. In a non-Bayesian framework, the tuning parameter is usually
chosen via cross-validation, but such a strategy ignores the uncertainty
about the choice of the tuning parameter.

Some studies consider Bayesian inference of two-component factorization,
$\boldsymbol{Y}=\boldsymbol{A}\boldsymbol{B}+\boldsymbol{U}$ (e.g.,
\citealp{Salakhutdinov2008,Babacan2012,Fazayeli2014}). These studies
assign independent normal priors to the elements of the unknown matrices.
This strategy corresponds to using the nuclear norm penalty or exponential
prior for the singular values, i.e., the diagonal elements of $\boldsymbol{\Omega}$,
in three-component factorization. In a similar vein to this study,
several studies propose Bayesian methods for inferring three-component
factorization (e.g., \citealp{Hoff2007,Zhou2010,Ding2011}). While
\citet{Hoff2007} use a normal prior for $\omega_{k}$ (in our notation),
\citet{Zhou2010} and \citet{Ding2011} model $\omega_{k}$ as a product
of two variables, $\omega_{k}=c_{k}^{2}d_{k}$, and use normal and
Beta-Bernoulli priors for $c_{k}$ and $d_{k}$, respectively. Therefore,
their prior can be regarded as a spike-and-slab version of the exponential
prior.

When $\pi_{k}=0$, $\forall k$, the proposed prior coincides with
the Lomax prior.\footnote{\citet{Tang2019} propose a positive generalized double Pareto prior,
but it is mathematically identical to a Lomax prior. } The hierarchical adaptive spectral penalty (HASP) \citep{Todeschini2013}
is seen as a non-Bayesian counterpart of the Lomax prior. The HASP
(Lomax prior) is more robust than the nuclear norm penalty (exponential
prior), in that the latter approach estimates the hyperparameter using
a Gamma prior. As suggested by the simulation studies in the next
section, in terms of the estimation accuracy, a Lomax prior is not
always better than an exponential prior.

Low-rank matrices are similar to static factor models (e.g., \citealp{Geweke1996,Arminger1998})
because in a factor model, a matrix of panel data is modeled as the
product of matrices of latent factors and factor loadings, that is,
a two-component factorization. As in the estimation of low-rank matrices,
the parameter identification problem is an important consideration
in the estimation of factor models. While some recent studies propose
approaches to address parameter identification in factor models (e.g.,
\citealp{Asmann2016,Chan2018,Fruehwirth-Schnatter2018}), these approaches
are limited to specific priors, and their prior specifications are
not specifically aimed at regularizing the complexity of the model.

\section{Simulation Study }

We compare the CSPE prior with alternative prior specifications by
means of a simulation study. The true matrix to be estimated $\boldsymbol{\Theta}_{0}$
is square, $J=T=30$, and its rank is $K^{*}\left(<K<J\right)$. $\boldsymbol{\Theta}_{0}$
is generated from the following static factor model:

\[
\boldsymbol{\Theta}_{0}=\boldsymbol{A}\boldsymbol{F},
\]

\[
\boldsymbol{F}=\left(\boldsymbol{f}_{1},...,\boldsymbol{f}_{J}\right),\quad\text{with}\;\boldsymbol{f}_{j}\sim\mathcal{N}\left(\boldsymbol{0}_{K^{*}},\boldsymbol{I}_{K^{*}}\right),
\]

\[
\boldsymbol{A}=\left(a_{j,k}\right),\quad\text{with}\;a_{j,k}\sim\mathcal{N}\left(0,1\right),
\]
where $\boldsymbol{F}$ is a $K^{*}\times J$ matrix composed of latent
factors and $\boldsymbol{A}$ is a $J\times K^{*}$ matrix of factor
loadings. $\boldsymbol{\Theta}_{0}$ is normalized to zero mean and
unit variance. $\boldsymbol{\Theta}_{0}$ is observed with additive
normal noise with variance $\tau^{-1}$; thus, the matrix of observations
$\boldsymbol{Y}$ is given by

\[
\boldsymbol{Y}=\boldsymbol{\Theta}_{0}+\boldsymbol{U},
\]
\[
\boldsymbol{U}=\left(u_{j,t}\right),\quad\text{with}\;u_{j,t}\sim\mathcal{N}\left(0,\tau^{-1}\right).
\]
$\tau$ is chosen such that the signal-to-noise ratio is 10. The dimension
of the model is fixed at $K=\left\lfloor J^{2}/\left(2J+1\right)\right\rfloor $,
where $\left\lfloor \right\rfloor $ denotes the floor function. Let
$M$ denote the number of missing entries. We consider three cases:
no missing, 10\% missing, and 90\% missing. These cases represent
estimation of complete matrices, imputation of missing observations
(or inference of treatment effects with panel data), and matrix completion,
respectively. Missing entries, if any, are chosen randomly.

We compare the proposed prior with four alternative prior specifications.
The first is a noninformative prior, $p\left(\omega_{k}\right)\propto\mathbb{I}\left(\omega_{k}\geq0\right)$,
$\forall k$. The second is an independent exponential prior, $\omega_{k}\sim\mathcal{E}\left(\chi\right)$,
$\forall k$, with $\chi\in\left\{ 1,0.5,0.1\right\} $. The third
is an independent Lomax prior, $\omega_{k}\sim\mathcal{L}\left(\mu_{1},\mu_{2}\right)$,
with $\mu_{1}=2$ and $\mu_{2}\in\left\{ 2,5,20\right\} $. The PDFs
of the alternatives are shown in Figure 3.

The fourth is a spike-and-slab exponential (SSE) prior specified by
the following hierarchy: 
\begin{eqnarray*}
\omega_{k}|\lambda_{k} & \sim & \mathcal{E}\left(\lambda_{k}\right),\quad k=1,...,K,\\
\lambda_{k}|\pi_{k} & \sim & \left(1-\pi_{k}\right)\mathcal{G}\left(\kappa_{1},\kappa_{2}\right)+\pi_{k}\delta,\quad k=1,...,K,\\
\pi_{k} & \sim & \mathcal{U}\left(0,1\right),\quad k=1,...,K,
\end{eqnarray*}
where $\mathcal{U}\left(a,b\right)$ denotes a uniform distribution
over an interval $\left(a,b\right)$. While a standard spike-and-slab
prior is defined by a mixture of two normal distributions with different
variances (e.g., \citealp{Ishwaran2005}), this prior is structured
as a mixture of two exponential distributions with different rates.
For the SSE prior, we use the same hyperparameters, $\delta$, $\kappa_{1}$,
and $\kappa_{2}$, as the CSPE prior. The SSE prior is different from
the CSPE prior in that the weights $\pi_{k}$ are independently and
uniformly distributed over a unit interval. Comparison between the
CSPE and SSE priors provides information about how the gains from
using the prior construction with a cumulative shrinkage process are
distinct from those from the spike-and-slab structure.

For all the alternatives, we fix $\nu_{1}=\nu_{2}=10^{-3}$ and $\bar{\eta}_{1}=\bar{\eta}_{2}=10^{\text{3}}$.
For each experiment, a total of 12,000 draws are simulated, and the
last 10,000 draws are used for estimation. The performance of the
alternative priors is evaluated in terms of the accuracy of the posterior
mean estimates of $\boldsymbol{\omega}$ and $\boldsymbol{\Theta}$
based on the sum of absolute errors (AE) and the sum of squared errors
(SE). Each measure is computed as the mean of 80 experiments.\footnote{We wrote all the programs in Matlab R2019b (64 bit) and executed them
on an Ubuntu Desktop 18.04 LTS (64 bit), running on AMD Ryzen Threadripper
1950X (4.2GHz).} 

Table 1 summarizes the results for the case with no missing values.
The CSPE prior is likely to outperform all the alternatives in terms
of estimation error in $\boldsymbol{\omega}$ and $\boldsymbol{\Theta}$,
especially when the true rank $K^{*}$ is smaller. As the SSE prior
falls between the exponential/Lomax priors and the CSPE prior, the
CSPE prior works well not only because of its spike-and-slab structure
but also the cumulative shrinkage applied to the weights. Although
when $K^{*}$ is smaller, the improvement in the estimation of $\boldsymbol{\omega}$
is likely to lead to more accurate estimates of $\boldsymbol{\Theta}$,
the CSPE prior is not significantly inferior to the others in terms
of AE/SE for $\boldsymbol{\Theta}$. The results for the 10\% missing
case are shown in Table 2, which yiels almost the same conclusions
as Table 1.

Table 3 reports the results for the 90\% missing case. As in the other
cases, the CSPE prior is competitive with the alternatives. Compared
to the noninformative prior, informative priors tend to have large
estimation errors for $\boldsymbol{\omega}$, but they are likely
to estimate $\boldsymbol{\Theta}$ more precisely. The CSPE prior
outperforms, or is at least comparable to, the other alternatives,
especially when the inferential problem is difficult, that is, the
true rank of $\boldsymbol{\Theta}$ is small and/or many data values
are missing.

As evident from Tables 1-3, the relationship between the relative
gains in the estimation of $\boldsymbol{\omega}$ and $\boldsymbol{\Theta}$
are not simple. A reduction in the estimation errors of $\boldsymbol{\omega}$
does not always imply that the estimation of $\boldsymbol{\Theta}$
is improved, and vice versa. Although the reason for such results
is not obvious, there is a possible explanation. As $\boldsymbol{\Phi}$
and $\boldsymbol{\Psi}$ are supposed to be unitary, the precision
in the estimation of $\boldsymbol{\Theta}$ largely depends on the
relative sizes of $\boldsymbol{\omega}$. Therefore, when insignificant
singular values are overestimated, even if the large singular values
are relatively precisely estimated, $\boldsymbol{\Theta}$ is inaccurately
estimated. In other words, for the sake of precise estimation of the
whole matrix $\boldsymbol{\Theta}$, it is more important to find
the true model, that is, the rank of $\boldsymbol{\Theta}$, than
to estimate the large singular values with high accuracy.

To investigate the difference between the alternatives in further
detail, we plot the relative accuracy of the posterior mean estimates
of the elements in $\boldsymbol{\omega}$ in terms of AE when $K^{*}=7$
and 10\% of entries are missing (Figure 4). The reported values are
normalized such that those for the noninformative prior are 100. Though
not reported here, similar patterns are observed for the other settings,
and the following discussion is not altered if the performance of
the alternative priors is measured based on SE.

As shown in Panel (a), for the exponential prior, irrespective of
$\chi$, the estimation errors of the large, significant singular
values are large, whereas the estimation errors of the nonsignificant
singular values are slightly smaller than those for the noninformative
prior. Thus, a trade-off exists between the relative accuracy in the
major and minor singular values. As $\chi$ increases, the estimation
errors for the large singular values decrease, while those for the
nonsignificant singular values increase.

Panel (b) of Figure 4 reports the results for the Lomax prior. Again,
the same trade-off is observed; however, compared to the results for
the exponential prior, the trade-off is mitigated to a certain degree,
as the estimation errors of the major singular values are comparatively
small.

The results for the SSE and CSPE priors are depicted in Panel (c).
While the estimation errors of the large singular values are comparable
to those for the Lomax prior, the estimation errors of the insignificant
singular values are smaller than those of the Lomax prior. Comparison
of the SSE and CSPE priors indicates that the estimation errors of
the significant singular values for these specifications are similar,
but when the CSPE prior is used, the nonnsignificant singular values
are likely to be estimated with higher precision. Therefore, the CSPE
prior effectively induces nonsignificant singular values to be approximately
zero and minimizes the errors in the estimation of large, significant
singular values.

Last, we compare the computational cost of the alternatives.Table
4 summarizes the computation time in seconds per 1,000 iterations,
which is measured using the \texttt{tic/toc} functions of Matlab.The
obtained computational times differ in a complicated manner depending
on the simulation environment because our posterior simulator is based
on NUTS, in which number of leap frog steps in each iteration depends
on the target kernel. When a prior density substantially complicates
the conditional posterior of $\boldsymbol{\vartheta}$, the number
of leap frog steps tends to be large. While the SSE and CSPE priors
have a disadvantage in that they involve additional Gibbs steps for
updating the hyperparameters, this approach does not appear to entail
a considerable computational burden.

\section{Application}

For illustration purposes, we apply the proposed approach to sectoral
industrial production in the United States (US). From the mid-1980s
to the mid-2000s, the volatility of major macroeconomic indicators,
such as real gross domestic product growth and price inflation rate,
decreased significantly in the US and other developed countries \citep{McConnell2000}.
Industrial production is not an exception; as shown in Figure 5, the
standard deviation of aggregate industrial production growth decreased
in the mid-1980s. In macroeconomics, this phenomenon is called the
Great Moderation, a term coined by \citealp{Stock2003} as a pun for
the Great Depression of 1929 to the mid-1930s. Many hypotheses for
the phenomenon, such as improved inventory management, good monetary
policy, and mere good luck (see, e.g., \citet{Summers2005,Davis2008}
for overview of the literature), have been proposed.

Although these hypotheses have been intensively examined empirically,
no consensus exists for the overall picture of the Great Moderation.
One unsettled issue is what happened before and after the Great Moderation
at a disaggregated level. In what follows, we investigate the changes
in the DGP of US industrial production during the Great Moderation
by means of Bayesian noisy matrix factorization.

The data used in this section are downloaded from the Federal Reserve
System's website.\footnote{https://www.federalreserve.gov/releases/g17/download.htm}.
Sectors are defined based on the International Standard Industrial
Classification Revision 4 (ISIC Rev. 4). The data matrices are three
panels of 25 sectors for 11 years (132 months). The time periods are
1973:1\textendash 1983:12, 1984:1\textendash 1994:12, and 1995:1\textendash 2005:12.
In what follows (including figures), when there is no fear of misunderstanding,
by omitting months, we denote the periods simply as 1973-1983, 1984-1994,
and 1995-2005, respectively. The start date is determined by the availability
of the data. Despite the fact that they use different statistical
methods, two notable early contributions, \citet{McConnell2000,Kim1999},
argue that the break date in the volatility of real GDP growth is
best estimated as 1984:Q1, which motivates the first splitting point.
The lengths of the second and third data are chosen such that all
the data span the same time period of 11 years (132 months). The data
are standardized by sector for each subsample.

Figure 6 includes scatter plots of the volatilities of sectoral industrial
production growth for different time periods. Plot (a) of Figure 6
compares 1973:1-1983:12 and 1984:1-1994:12. Although, as seen in Figure
5, the aggregate volatility decreased during these periods, the volatilities
in 20 (of 25) sectors increased. In turn, plot (b) of Figure 6 compares
1984:1-1994:12 and 1995:1-2005:12. The number of sectors that experienced
an increase in volatility during this period is 12. Therefore, sectoral
data appear to offer a different picture of the Great Moderation than
aggregate data.

We infer a three-component matrix factorization of a matrix containing
the sectoral industrial production growth for different periods:

\begin{eqnarray*}
\boldsymbol{Y} & = & \boldsymbol{\Theta}+\boldsymbol{U}\\
 & = & \boldsymbol{\Phi}\boldsymbol{\Omega}\boldsymbol{\Psi}^{\top}+\boldsymbol{U},
\end{eqnarray*}
\[
\boldsymbol{U}=\left(u_{j,t}\right),\quad\text{with}\;u_{j,t}\sim\mathcal{N}\left(0,\tau^{-1}\right).
\]
With $J=25$ and $T=132$, we choose $K=\left\lfloor J^{2}/\left(2J+1\right)\right\rfloor =20$.
The hyperparameters are set to the same values as in Section 3. In
what follows, we report only the results for the conservative choice
of $\alpha$: when the aggressive choice is used, the conclusions
are virtually the same. A total of 420,000 posterior draws are simulated,
and the last 400,000 draws are used for posterior analysis.

An estimated singular value decomposition has no explicit structural
interpretation, but it conveys information about the underlying DGP.
An estimated singular value decomposition has close ties to a static
factor model, and the model is represented as

\[
\boldsymbol{Y}=\sum_{k=1}^{K}\omega_{k}\boldsymbol{\phi}_{k}\boldsymbol{\psi}_{k}^{\top}+\boldsymbol{U}.
\]
As $\boldsymbol{\Phi}$ and $\boldsymbol{\Psi}$ are supposed to be
unitary, the $k$th element of $\boldsymbol{\omega}$ can be interpreted
as the magnitude of the $k$th ``latent factor'', and its contribution
to the overall variance is treated as independent from the other latent
factors and errors. The variance of the observed data can be decomposed
as 
\begin{eqnarray*}
\mathbb{V}\left[y_{j,t}\right] & = & \sum_{k=1}^{K}\mathbb{V}\left[\omega_{k}\phi_{j,k}\psi_{t,k}\right]+\mathbb{V}\left[u_{j,t}\right]\\
 & = & \sum_{k=1}^{K}\omega_{k}^{2}+\tau^{-1}.
\end{eqnarray*}
Note that the data are standardized by sector for each subsample;
thus, all the variances of the subsamples are the same, $\mathbb{V}\left[y_{j,t}\right]=1$.
The contributions of $\omega_{k}^{2}$ and $\tau^{-1}$ to the variance
of $y_{j,t}$ are interpreted in relative terms. Our primary interest
is to investigate the difference in the relative contributions of
the singular values and the errors for different time periods. $\omega_{k}^{2}$
for small $k$ can be interpreted as the magnitude of shocks that
affect broad sectors, while $\omega_{k}^{2}$ for large $k$ can be
seen as the magnitude of shocks that affect a portion of sectors.
$\tau^{-1}$ can viewed as the magnitude of idiosyncratic (sector-specific)
shocks that do not spillover to other sectors.

Several studies report that the decrease in the volatility of macroeconomic
indicators is partly caused by decreases in comovement between sectors
or firms, e.g., \citet{Comin2006,Irvine2007,Stiroh2009,Burren2013}.
While these studies differ in terms of data and method, they all adopt
non-Bayesian approaches; thus, they do not provide any uncertainty
quantification.

Among the literature, \citet{Foerster2011} is the most closely related
to our study in that they estimate a static approximate factor model
\citep{Chamberlain1983} for data on US sectoral industrial production.
They report that nearly all the variability is associated with common
factors; thus, the decrease in the volatility of common shocks largely
explains the Great Moderation. They also argue that the relative importance
of sector-specific shocks increased during the Great Moderation period.
Their analysis relies on a non-Bayesian method and does not evaluate
the uncertainty about the analysis. As their approximate factor model
has only two factors, it is in danger of distorting the overall picture.
Our framework can accommodate a large number of factors (singular
values) and can serve as a robustness check of the results of \citet{Foerster2011}.

Figure 7 plots the posterior estimates of $\boldsymbol{\omega}$.
All the panels include the same lines representing the posterior mean
estimates of $\boldsymbol{\omega}$ for different periods. The shaded
areas are the corresponding 95\% credible sets for each period. There
are two results worth noting. First, $\omega_{k}$s with small $k$
(largest singular values) for the periods after 1984 are estimated
to be smaller than those for the period before 1984, which implies
that the variability of shocks that impact many sectors decreased
during the Great Moderation period. Second, the posterior mean estimates
of $\omega_{k}$s with larger $k$ (less important singular values)
for the Great Moderation period are smaller than those for the period
before the Great Moderation. Thus, shocks that affect only a portion
of sectors became more important during the Great Moderation. These
results coincide with the abovementioned studies that report a decrease
in comovement between sectors.

Figure 8 plots the posterior density of the error variance $\tau^{-1}$
for each period. The posterior density of $\tau^{-1}$ for 1973-1983
is significantly smaller than that for 1984-1994. The posterior means
of the signal-to-noise ratio for 1973-1983, 1984-1994, and 1995-2005
are 14.3, 5.2, and 6.8, respectively, which indicates that the relative
contribution of idiosyncratic shocks to the variability of sectoral
industrial production decreased during the Great Moderation.

Figure 9 shows the differences in the posterior mean estimates of
$\omega_{k}^{2}$ and $\tau^{-1}$ for consecutive periods. Panel
(a) plots the differences between 1973-1983 and 1984-1994. The largest
difference is found in $\omega_{1}$ (largest singular value), followed
by that in $\tau^{-1}$. Our results are consistent with those of
\citet{Foerster2011}. Their approximate factor model has only two
latent factors and the remaining less important factors are treated
as idiosyncratic shocks. Our results suggest that although \citet{Foerster2011}
overestimated the increases in the relative importance of idiosyncratic
shocks, their conclusion is largely robust. Panel (b) of Figure 9
plots the differences in the posterior estimates of $\omega_{k}^{2}$
and $\tau^{-1}$ between 1984-1994 and 1995-2005. Modest reversions
in $\omega_{1}^{2}$ and $\tau^{-1}$ are observed, but there is no
salient difference for the periods.

\section{Concluding Remarks}

This study proposes a novel prior specification for low-rank matrix
factorization called the cumulative shrinkage process exponential
(CSPE) prior. The proposed prior is specified by a scale mixture of
exponential distributions with spike and slab components. The weights
for the spike/slab parts are inferred via a tailored prior based on
a cumulative shrinkage process \citep{Legramantiforthcoming}. The
proposed prior increasingly aggressively shrinks less important singular
values toward zero. By means of a simulation study, we show that our
proposal is competitive with other alternative priors. For illustration,
we apply the proposed approach to the US sectoral industrial production
to analyze the Great Moderation.

We note two research topics to be studied in the future. First, it
would be desirable to develop a more systematic procedure to choose
the hyperparameters for the CSPE prior. Although we describe an elicitation
strategy, there is room for further development. Second, though not
specific to matrix factorization, more efficient methods to sample
from constrained spaces are sorely needed. In particular, sampling
from a Stiefel manifold (more generally, orthogonal spaces) is challenging,
and there are a number of applications that involve constraints of
this kind.

\section*{Disclosure Statement}

The author declares that there is no conflict of interest regarding
the publication of this paper. This research received no specific
grant from any funding agency in the public, commercial, or not-for-profit
sectors. 

\bibliographystyle{econ}
\bibliography{reference}

\section*{Appendix}

\paragraph{Posterior simulator}

A single cycle of the posterior simulator is summarized in the following.
\begin{enumerate}
\item Update $\eta_{l}$:

\[
\eta_{l}=\begin{cases}
\bar{\eta}_{l}\left[1-\exp\left(\frac{\log\left(1-a_{\eta}\right)}{b_{\eta}\bar{\eta}_{l}}\times i\right)\right], & i<\bar{\eta}_{l}\\
\bar{\eta}_{l}, & i\geq\bar{\eta}_{l}
\end{cases},\quad l=1,2.
\]

\item Sample $\boldsymbol{\vartheta}$:

Sample $\boldsymbol{\vartheta}$ using the No-U-Turn sampler (Hoffman
and Gelman, 2014).
\item Sample $z_{k}$:

\[
p\left(z_{k}=l|\text{rest}\right)\propto\begin{cases}
\pi_{l}f_{E}\left(\omega_{k}|\delta\right), & l=1,...,k,\\
\pi_{l}f_{L}\left(\omega_{k}|\kappa_{1},\kappa_{2}\right), & l=k+1,...,K.
\end{cases}
\]

\item Sample $\upsilon_{k}$:

\[
\upsilon_{k}|\text{rest}\sim\mathcal{B}\left(1+\sum_{l=1}^{K}\mathbb{I}\left(z_{l}=k\right),\;\alpha+\sum_{l=1}^{K}\mathbb{I}\left(z_{l}>k\right)\right),\quad k=1,...,K-1.
\]

\item Sample $\lambda_{k}$:

\[
\lambda_{k}|\text{rest}\sim\mathbb{I}\left(z_{k}\leq k\right)\delta_{\infty}+\left(1-\mathbb{I}\left(z_{k}\leq k\right)\right)\mathcal{G}\left(\kappa_{1}+1,\;\kappa_{2}+\omega_{k}\right),\quad k=1,...,K.
\]

\item Sample $\tau$:

\[
\tau|\text{rest}\sim\mathcal{G}\left(\nu_{1}+\frac{JT}{2},\;\nu_{2}+\frac{1}{2}\textrm{vec}\left(\boldsymbol{Y}-\boldsymbol{\Theta}\right)^{\top}\textrm{vec}\left(\boldsymbol{Y}-\boldsymbol{\Theta}\right)\right).
\]

\item Sample the missing elements of $\boldsymbol{Y}$ (if any):

\[
y_{j,t}|\text{rest}\sim\mathcal{N}\left(\theta_{j,t},\tau^{-1}\right),\quad\left(j,t\right)\in\mathcal{I}_{miss}.
\]

\end{enumerate}

\paragraph{The gradient of $\log p\left(\boldsymbol{\vartheta}|\mathrm{rest}\right)$}

The gradient of the log approximate conditional posterior of $\boldsymbol{\vartheta}$
is computed as follows: 
\[
\nabla_{\boldsymbol{\vartheta}}\log p\left(\boldsymbol{\vartheta}|\text{rest}\right)=\left(\begin{array}{c}
\nabla_{\textrm{vec}\left(\boldsymbol{\Phi}\right)}\log p\left(\boldsymbol{\vartheta}|\text{rest}\right)\\
\nabla_{\textrm{vec}\left(\boldsymbol{\Psi}\right)}\log p\left(\boldsymbol{\vartheta}|\text{rest}\right)\\
\nabla_{\boldsymbol{\omega}^{*}}\log p\left(\boldsymbol{\vartheta}|\text{rest}\right)
\end{array}\right),
\]
\[
\nabla_{\textrm{vec}\left(\boldsymbol{\Phi}\right)}\log p\left(\boldsymbol{\vartheta}|\text{rest}\right)=-\tau\textrm{vec}\left(-\boldsymbol{Y}\boldsymbol{\Psi}\boldsymbol{\Omega}+\boldsymbol{\Phi}\boldsymbol{\Omega}\boldsymbol{\Psi}^{\top}\boldsymbol{\Psi}\boldsymbol{\Omega}\right)-4\eta_{1}\textrm{vec}\left(\boldsymbol{\Phi}\boldsymbol{\Phi}^{\top}\boldsymbol{\Phi}-\boldsymbol{\Phi}\right),
\]
\[
\nabla_{\textrm{vec}\left(\boldsymbol{\Psi}\right)}\log p\left(\boldsymbol{\vartheta}|\text{rest}\right)=-\tau\textrm{vec}\left(-\boldsymbol{Y}^{\top}\boldsymbol{\Phi}\boldsymbol{\Omega}+\boldsymbol{\Psi}\boldsymbol{\Omega}\boldsymbol{\Phi}^{\top}\boldsymbol{\Phi}\boldsymbol{\Omega}\right)-4\eta_{1}\textrm{vec}\left(\boldsymbol{\Psi}\boldsymbol{\Psi}^{\top}\boldsymbol{\Psi}-\boldsymbol{\Psi}\right),
\]
\[
\nabla_{\boldsymbol{\omega}^{*}}\log p\left(\boldsymbol{\vartheta}|\text{rest}\right)=-\tau\left(-\boldsymbol{\Upsilon}^{\top}\textrm{vec}\left(\boldsymbol{Y}\right)+\boldsymbol{\Upsilon}^{\top}\boldsymbol{\Upsilon}\boldsymbol{\omega}^{*}\right)-\left(\boldsymbol{C}^{-1}\right)^{\top}\boldsymbol{\lambda}+\nabla_{\boldsymbol{\omega}^{*}}\rho\left(\boldsymbol{\omega}^{*}\right),
\]
\[
\boldsymbol{\Upsilon}=\left(\textrm{vec}\left(\boldsymbol{\phi}_{1}\boldsymbol{\psi}_{1}^{\top}\right),...,\textrm{vec}\left(\boldsymbol{\phi}_{K}\boldsymbol{\psi}_{K}^{\top}\right)\right)\boldsymbol{C}^{-1},
\]
\[
\nabla_{\boldsymbol{\omega}^{*}}\rho\left(\boldsymbol{\omega}^{*}\right)=\left(\nabla_{\omega_{1}^{*}}\rho\left(\omega_{1}^{*}\right),...,\nabla_{\omega_{K}^{*}}\rho\left(\omega_{K}^{*}\right)\right)^{\top},
\]
\[
\nabla_{\omega_{k}^{*}}\rho\left(\omega_{k}^{*}\right)=\frac{\eta_{2}\exp\left(-\eta_{2}\omega_{k}^{*}\right)}{1+\exp\left(-\eta_{2}\omega_{k}^{*}\right)},\quad k=1,...,K.
\]

\clearpage{}

\begin{sidewaystable}
\caption{Results of Monte Carlo experiment (1): No missing}

\medskip{}

\begin{centering}
\begin{tabular}{llcccccccccccccc}
\hline 
Prior  & \multicolumn{1}{l}{} & \multicolumn{4}{c}{$K^{*}=12$} &  & \multicolumn{4}{c}{$K^{*}=7$} &  & \multicolumn{4}{c}{$K^{*}=3$}\tabularnewline
 &  & \multicolumn{2}{c}{$\boldsymbol{\omega}$} & \multicolumn{2}{c}{$\boldsymbol{\Theta}$} &  & \multicolumn{2}{c}{$\boldsymbol{\omega}$} & \multicolumn{2}{c}{$\boldsymbol{\Theta}$} &  & \multicolumn{2}{c}{$\boldsymbol{\omega}$} & \multicolumn{2}{c}{$\boldsymbol{\Theta}$}\tabularnewline
 &  & AE  & SE  & AE  & SE  &  & AE  & SE  & AE  & SE  &  & AE  & SE  & AE  & SE\tabularnewline
\hline 
Noninformative  &  & 100.0  & 100.0  & 100.0  & 100.0  &  & 100.0  & 100.0  & 100.0  & 100.0  &  & 100.0  & 100.0  & 100.0  & 100.0\tabularnewline
\hline 
Exponential  & $\chi=1$  & 102.2  & 95.2 & 99.9  & 99.9  &  & 94.7  & 86.3  & 99.2  & 98.4  &  & 91.4  & 84.3  & 96.5  & 93.6\tabularnewline
 & $\chi=0.5$  & 100.9  & 97.5  & 100.0  & 99.9  &  & 97.0  & 92.8  & 99.5  & 99.1  &  & 95.9  & 92.6  & 98.4  & 96.9\tabularnewline
 & $\chi=0.1$  & 99.7  & 98.5  & 100.0  & 99.9  &  & 99.2  & 98.4  & 99.9  & 99.8  &  & 99.0  & 98.3  & 99.6  & 99.3\tabularnewline
\hline 
Lomax  & $\mu_{2}=2$  & 96.2  & 87.2  & 99.9  & 99.8  &  & 89.2  & 80.6  & 99.0  & 98.1  &  & 85.5  & 78.5  & 95.9  & 92.3\tabularnewline
 & $\mu_{2}=5$  & 98.6  & 94.3  & 100.0  & 99.9  &  & 95.4  & 91.3  & 99.5  & 99.1  &  & 93.5  & 90.0  & 98.0  & 96.2\tabularnewline
 & $\mu_{2}=20$  & 99.9  & 98.5  & 100.0  & 99.9  &  & 99.2  & 98.3  & 99.9  & 99.8  &  & 98.2  & 97.2  & 99.4  & 98.9\tabularnewline
\hline 
SSE  &  & 85.4  & 65.9  & \textbf{99.9}  & \textbf{99.7}  &  & 87.6  & 79.9  & 98.9  & 98.0  &  & 89.1  & 84.6  & 97.0  & 94.3\tabularnewline
\hline 
CSPE  & conservative  & 82.5  & 59.6  & 99.9  & 99.9  &  & 80.4  & 70.4  & 98.6  & 97.2  &  & 80.5  & 75.2  & 95.3  & 91.3\tabularnewline
 & aggressive  & \textbf{79.7}  & \textbf{53.0}  & 100.0  & 100.0  &  & \textbf{74.0}  & \textbf{60.7}  & \textbf{98.2}  & \textbf{96.6}  &  & \textbf{74.6}  & \textbf{67.9}  & \textbf{94.2}  & \textbf{89.4}\tabularnewline
\hline 
\end{tabular}
\par\end{centering}
\begin{centering}
\medskip{}
 
\par\end{centering}
\centering{}%
\begin{minipage}[t]{0.8\columnwidth}%
Note: SE denotes the median of the sum of the squared errors. AE denotes
the median of the sum of the absolute errors. Reported values are
normalized by the corresponding values for the noninformative prior. %
\end{minipage}
\end{sidewaystable}

\clearpage{}

\begin{sidewaystable}
\caption{Results of Monte Carlo experiment (2): 10\% missing}

\medskip{}

\begin{centering}
\begin{tabular}{llcccccccccccccc}
\hline 
Prior  & \multicolumn{1}{l}{} & \multicolumn{4}{c}{$K^{*}=12$} &  & \multicolumn{4}{c}{$K^{*}=7$} &  & \multicolumn{4}{c}{$K^{*}=3$}\tabularnewline
 &  & \multicolumn{2}{c}{$\boldsymbol{\omega}$} & \multicolumn{2}{c}{$\boldsymbol{\Theta}$} &  & \multicolumn{2}{c}{$\boldsymbol{\omega}$} & \multicolumn{2}{c}{$\boldsymbol{\Theta}$} &  & \multicolumn{2}{c}{$\boldsymbol{\omega}$} & \multicolumn{2}{c}{$\boldsymbol{\Theta}$}\tabularnewline
 &  & AE  & SE  & AE  & SE  &  & AE  & SE  & AE  & SE  &  & AE  & SE  & AE  & SE\tabularnewline
\hline 
Noninformative  &  & 100.0  & 100.0  & 100.0  & 100.0  &  & 100.0  & 100.0  & 100.0  & 100.0  &  & 100.0  & 100.0  & 100.0  & 100.0\tabularnewline
\hline 
Exponential  & $\chi=1$  & 99.5  & 91.7  & 100.0  & 100.0  &  & 94.6  & 86.1  & 99.1  & 98.0  &  & 90.8  & 83.5  & 96.5  & 93.6\tabularnewline
 & $\chi=0.5$  & 99.0  & 93.8  & 100.0  & 99.9  &  & 97.2  & 93.0  & 99.6  & 99.0  &  & 95.1  & 91.1  & 98.1  & 96.5\tabularnewline
 & $\chi=0.1$  & 100.1  & 100.0  & 100.1  & 100.0  &  & 99.4  & 98.6  & 99.9  & 99.8  &  & 98.9  & 98.1  & 99.6  & 99.2\tabularnewline
\hline 
Lomax  & $\mu_{2}=2$  & 95.1  & 84.5  & 100.0  & 99.9  &  & 88.4  & 79.2  & 98.9  & 97.7  &  & 85.1  & 77.5  & 95.9  & 92.5\tabularnewline
 & $\mu_{2}=5$  & 97.6  & 92.4  & 100.0  & 100.0  &  & 95.4  & 91.4  & 99.5  & 98.8  &  & 93.4  & 89.7  & 98.0  & 96.3\tabularnewline
 & $\mu_{2}=20$  & 99.8  & 98.8  & 100.1  & 100.2  &  & 98.8  & 97.7  & 99.8  & 99.6  &  & 98.4  & 97.1  & 99.3  & 98.7\tabularnewline
\hline 
SSE  &  & 83.7  & 62.5  & \textbf{100.0}  & \textbf{99.8}  &  & 87.7  & 79.9  & 98.9  & 97.7  &  & 88.6  & 84.1  & 96.9  & 94.3\tabularnewline
\hline 
CSPE  & conservative  & 81.3  & 56.6  & 100.0  & 100.1  &  & 79.9  & 69.5  & 98.5  & 96.8  &  & 79.7  & 73.9  & 95.3  & 91.4\tabularnewline
 & aggressive  & \textbf{79.1}  & \textbf{52.1}  & 100.2  & 100.4  &  & \textbf{73.9}  & \textbf{60.7}  & \textbf{98.2}  & \textbf{96.1}  &  & \textbf{73.5}  & \textbf{66.1}  & \textbf{94.1}  & \textbf{89.4}\tabularnewline
\hline 
\end{tabular}
\par\end{centering}
\begin{centering}
\medskip{}
 
\par\end{centering}
\centering{}%
\begin{minipage}[t]{0.8\columnwidth}%
Note: SE denotes the median of the sum of the squared errors. AE denotes
the median of the sum of the absolute errors. Reported values are
normalized by the corresponding values for the noninformative prior. %
\end{minipage}
\end{sidewaystable}

\clearpage{}

\begin{sidewaystable}
\caption{Result of Monte Carlo experiment (3): 90\% missing}

\medskip{}

\begin{centering}
\begin{tabular}{llcccccccccccccc}
\hline 
Prior  & \multicolumn{1}{l}{} & \multicolumn{4}{c}{$K^{*}=12$} &  & \multicolumn{4}{c}{$K^{*}=7$} &  & \multicolumn{4}{c}{$K^{*}=3$}\tabularnewline
 &  & \multicolumn{2}{c}{$\boldsymbol{\omega}$} & \multicolumn{2}{c}{$\boldsymbol{\Theta}$} &  & \multicolumn{2}{c}{$\boldsymbol{\omega}$} & \multicolumn{2}{c}{$\boldsymbol{\Theta}$} &  & \multicolumn{2}{c}{$\boldsymbol{\omega}$} & \multicolumn{2}{c}{$\boldsymbol{\Theta}$}\tabularnewline
 &  & AE  & SE  & AE  & SE  &  & AE  & SE  & AE  & SE  &  & AE  & SE  & AE  & SE\tabularnewline
\hline 
Noninformative  &  & 100.0  & 100.0  & 100.0  & 100.0  &  & 100.0  & 100.0  & 100.0  & 100.0  &  & 100.0  & 100.0  & 100.0  & 100.0\tabularnewline
\hline 
Exponential  & $\chi=1$  & 552.2  & 3075.8  & 99.4  & 91.7  &  & 221.4  & 660.5  & 98.6  & 90.8  &  & 84.4  & 184.1  & 96.5  & 88.8\tabularnewline
 & $\chi=0.5$  & 399.0  & 1577.4  & 97.9  & 90.3  &  & 158.5  & 304.5  & 97.0  & 89.9  &  & 71.6  & 73.0  & 93.9  & 88.7\tabularnewline
 & $\chi=0.1$  & \textbf{110.5}  & \textbf{130.2}  & 99.2  & 98.9  &  & 96.5  & 94.8  & 99.4  & 100.3  &  & 91.8  & 89.1  & 98.5  & 98.7\tabularnewline
\hline 
Lomax  & $\mu_{2}=2$  & 403.2  & 1660.0  & 98.6  & 93.0  &  & 141.8  & 269.3  & 97.7  & 92.3  &  & 53.3  & 47.1  & 95.9  & 95.5\tabularnewline
 & $\mu_{2}=5$  & 254.4  & 626.8  & 98.4  & 95.5  &  & 102.4  & 107.2  & 96.8  & 93.3  &  & 69.1  & 57.1  & 96.4  & 96.3\tabularnewline
 & $\mu_{2}=20$  & 117.8  & 160.6  & 98.9  & 98.7  &  & 91.6  & 83.7  & 97.0  & 94.9  &  & 88.2  & 85.1  & 98.5  & 99.8\tabularnewline
\hline 
SSE  &  & 152.2  & 247.9  & 96.5  & 93.3  &  & \textbf{79.6}  & \textbf{67.4}  & 97.3  & 95.7  &  & 79.2  & 73.2  & 97.2  & 96.6\tabularnewline
\hline 
CSPE  & conservative  & 311.3  & 908.2  & \textbf{96.3}  & 89.6  &  & 102.3  & 128.0  & 95.1  & 89.2  &  & 51.3  & 40.1  & \textbf{94.2}  & \textbf{91.8}\tabularnewline
 & aggressive  & 368.5  & 1265.7  & 96.7  & \textbf{89.3}  &  & 118.0  & 182.8  & \textbf{95.1}  & \textbf{87.9}  &  & \textbf{45.1}  & \textbf{36.8}  & 95.4  & 94.3\tabularnewline
\hline 
\end{tabular}
\par\end{centering}
\begin{centering}
\medskip{}
 
\par\end{centering}
\centering{}%
\begin{minipage}[t]{0.8\columnwidth}%
Note: SE denotes the median of the sum of the squared errors. AE denotes
the median of the sum of the absolute errors. Reported values are
normalized by the corresponding values for the noninformative prior. %
\end{minipage}
\end{sidewaystable}

\clearpage{}

\begin{table}
\caption{Computation time}

\medskip{}

\begin{centering}
\begin{tabular}{llr@{\extracolsep{0pt}.}lr@{\extracolsep{0pt}.}lr@{\extracolsep{0pt}.}lr@{\extracolsep{0pt}.}lr@{\extracolsep{0pt}.}lr@{\extracolsep{0pt}.}lr@{\extracolsep{0pt}.}lr@{\extracolsep{0pt}.}lr@{\extracolsep{0pt}.}l}
\hline 
 &  & \multicolumn{6}{c}{No missing} & \multicolumn{6}{c}{10\% missing} & \multicolumn{6}{c}{90\% missing}\tabularnewline
$K^{*}$  &  & \multicolumn{2}{c}{12} & \multicolumn{2}{c}{7} & \multicolumn{2}{c}{3} & \multicolumn{2}{c}{12} & \multicolumn{2}{c}{7} & \multicolumn{2}{c}{3} & \multicolumn{2}{c}{12} & \multicolumn{2}{c}{7} & \multicolumn{2}{c}{3}\tabularnewline
\hline 
Noninformative  &  & 45 & 5  & 54 & 6  & 61 & 9  & 46 & 4  & 56 & 1  & 63 & 8  & 73 & 4  & 70 & 0  & 68 & 1\tabularnewline
\hline 
Exponential  & $\chi=1$  & 48 & 4  & 56 & 3  & 59 & 6  & 48 & 2  & 54 & 8  & 82 & 0  & 50 & 5  & 50 & 7  & 49 & 4\tabularnewline
 & $\chi=0.5$  & 48 & 1  & 54 & 0  & 61 & 4  & 50 & 0  & 55 & 7  & 64 & 7  & 48 & 7  & 49 & 0  & 54 & 1\tabularnewline
 & $\chi=0.1$  & 48 & 8  & 56 & 1  & 64 & 4  & 48 & 3  & 55 & 6  & 62 & 7  & 73 & 2  & 70 & 8  & 69 & 8\tabularnewline
\hline 
Lomax  & $\mu_{2}=2$  & 48 & 5  & 53 & 7  & 61 & 2  & 48 & 3  & 55 & 4  & 61 & 8  & 49 & 6  & 50 & 2  & 51 & 1\tabularnewline
 & $\mu_{2}=5$  & 47 & 1  & 54 & 7  & 61 & 9  & 48 & 5  & 54 & 7  & 61 & 4  & 53 & 9  & 55 & 1  & 59 & 4\tabularnewline
 & $\mu_{2}=20$  & 48 & 3  & 56 & 0  & 64 & 6  & 48 & 3  & 54 & 6  & 60 & 8  & 68 & 8  & 68 & 2  & 67 & 3\tabularnewline
\hline 
SSE  &  & 51 & 5  & 57 & 7  & 60 & 4  & 50 & 8  & 54 & 3  & 61 & 9  & 59 & 7  & 59 & 5  & 63 & 8\tabularnewline
\hline 
CSPE  & conservative  & 52 & 0  & 56 & 8  & 62 & 8  & 52 & 6  & 56 & 5  & 61 & 3  & 55 & 0  & 56 & 8  & 57 & 0\tabularnewline
 & aggressive  & 52 & 7  & 59 & 3  & 63 & 2  & 54 & 5  & 58 & 8  & 62 & 6  & 55 & 5  & 58 & 2  & 59 & 7\tabularnewline
\hline 
\end{tabular}
\par\end{centering}
\centering{}%
\begin{minipage}[t]{0.8\columnwidth}%
Note: Numbers indicate the mean computation time per 1,000 iterations
measured in seconds.%
\end{minipage}
\end{table}
0

\clearpage{}

\begin{figure}
\caption{PDFs of the spike and slab parts of the CSPE prior}

\centering{}\includegraphics{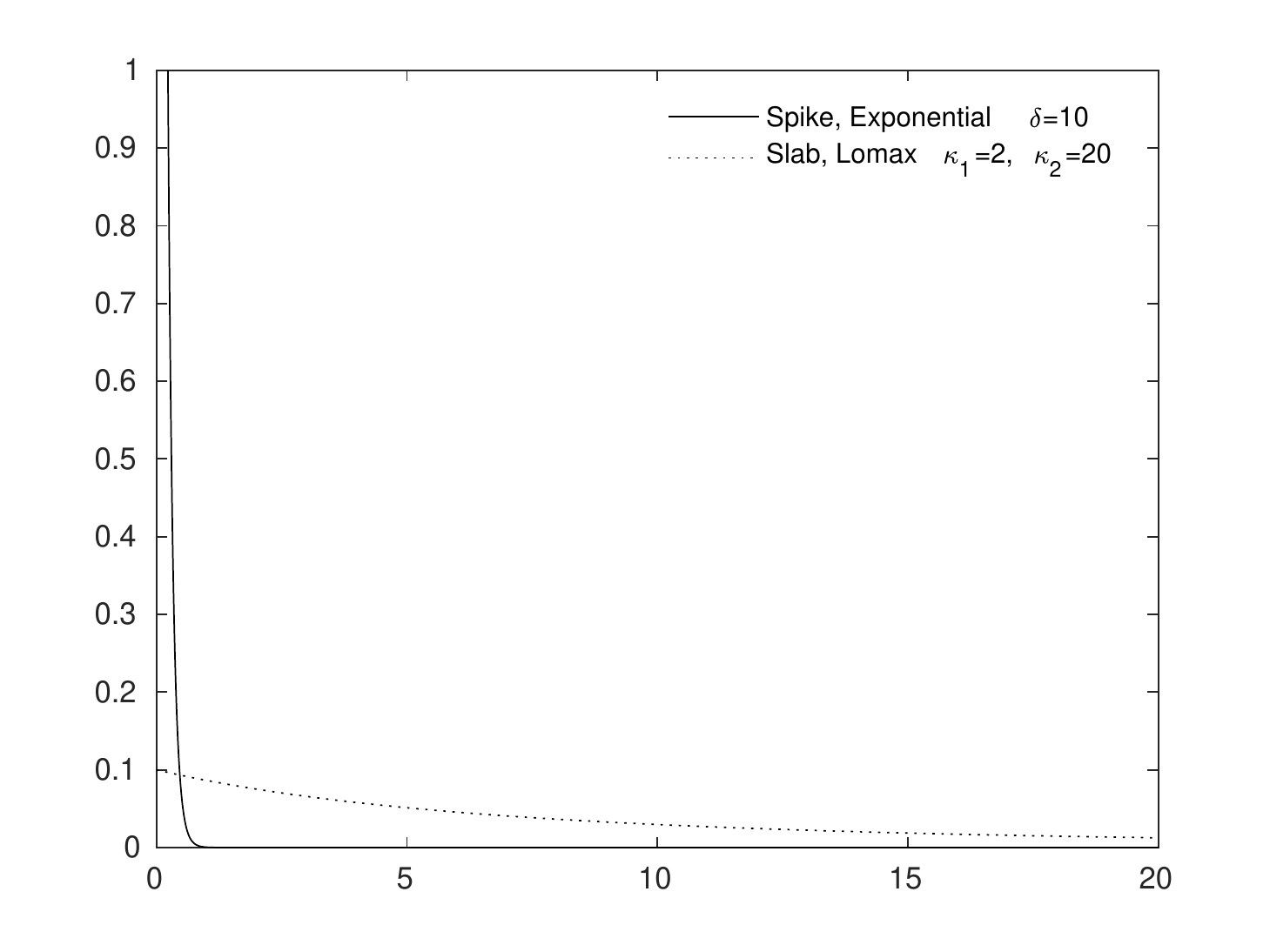} 
\end{figure}

\clearpage{}

\begin{figure}
\caption{Prior means of the weights on the spike and slab parts of the CSPE
prior}

\begin{centering}
\includegraphics[scale=1.2]{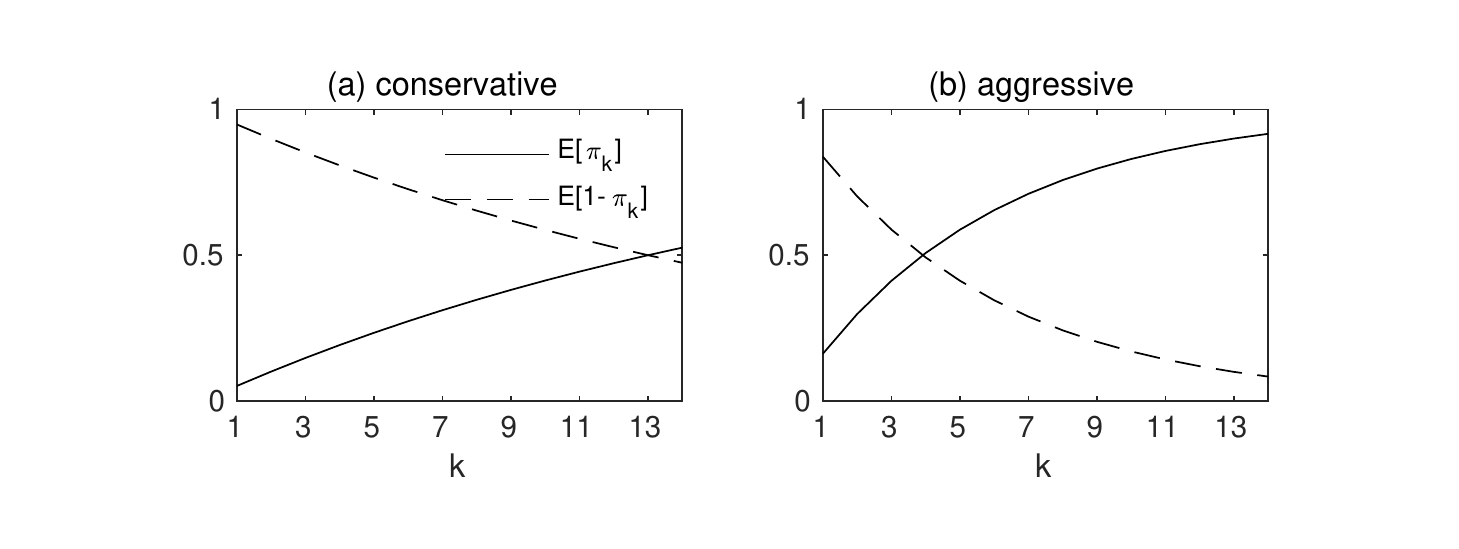} 
\par\end{centering}
\centering{}%
\begin{minipage}[t]{0.8\columnwidth}%
Note: $K=14$. For conservative and aggressive cases, $\alpha$ is
chosen such that the probability of the $K-1$th singular value being
insignificant is 0.5 or 0.9, respectively. %
\end{minipage}
\end{figure}

\clearpage{}

\begin{figure}
\caption{PDFs of the exponential and Lomax priors}

\centering{}\includegraphics[scale=1.2]{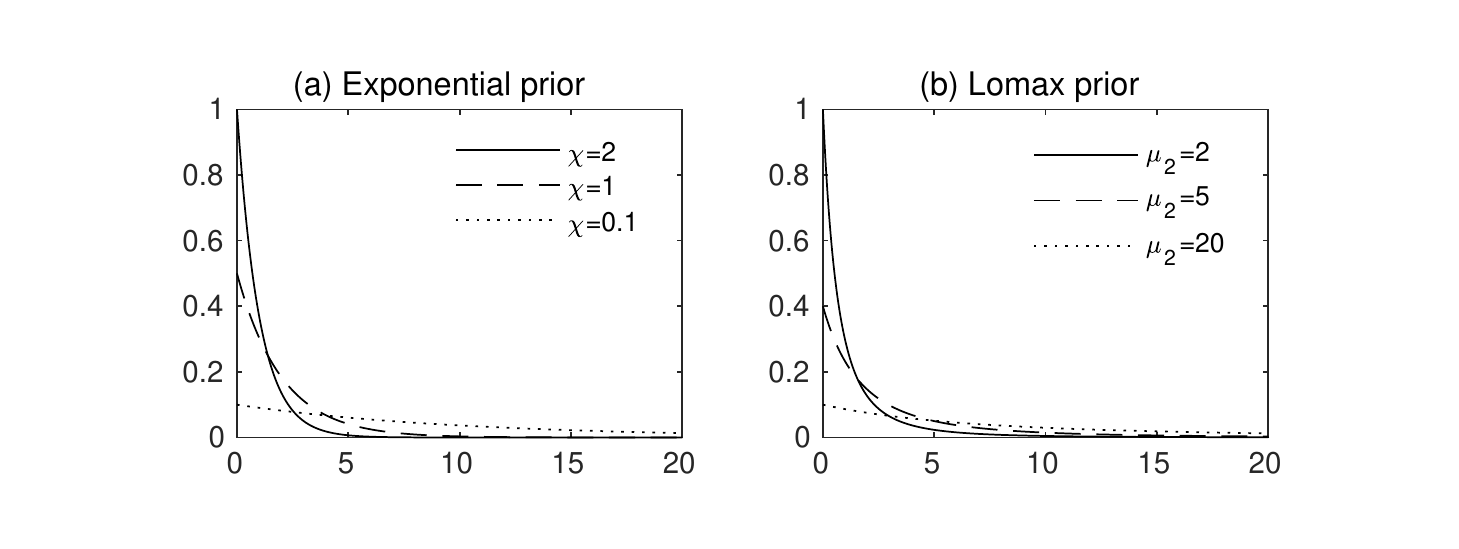} 
\end{figure}

\clearpage{}

\begin{figure}
\caption{Sum of the absolute errors of $\boldsymbol{\omega}$}

\centering{}\includegraphics[scale=1.2]{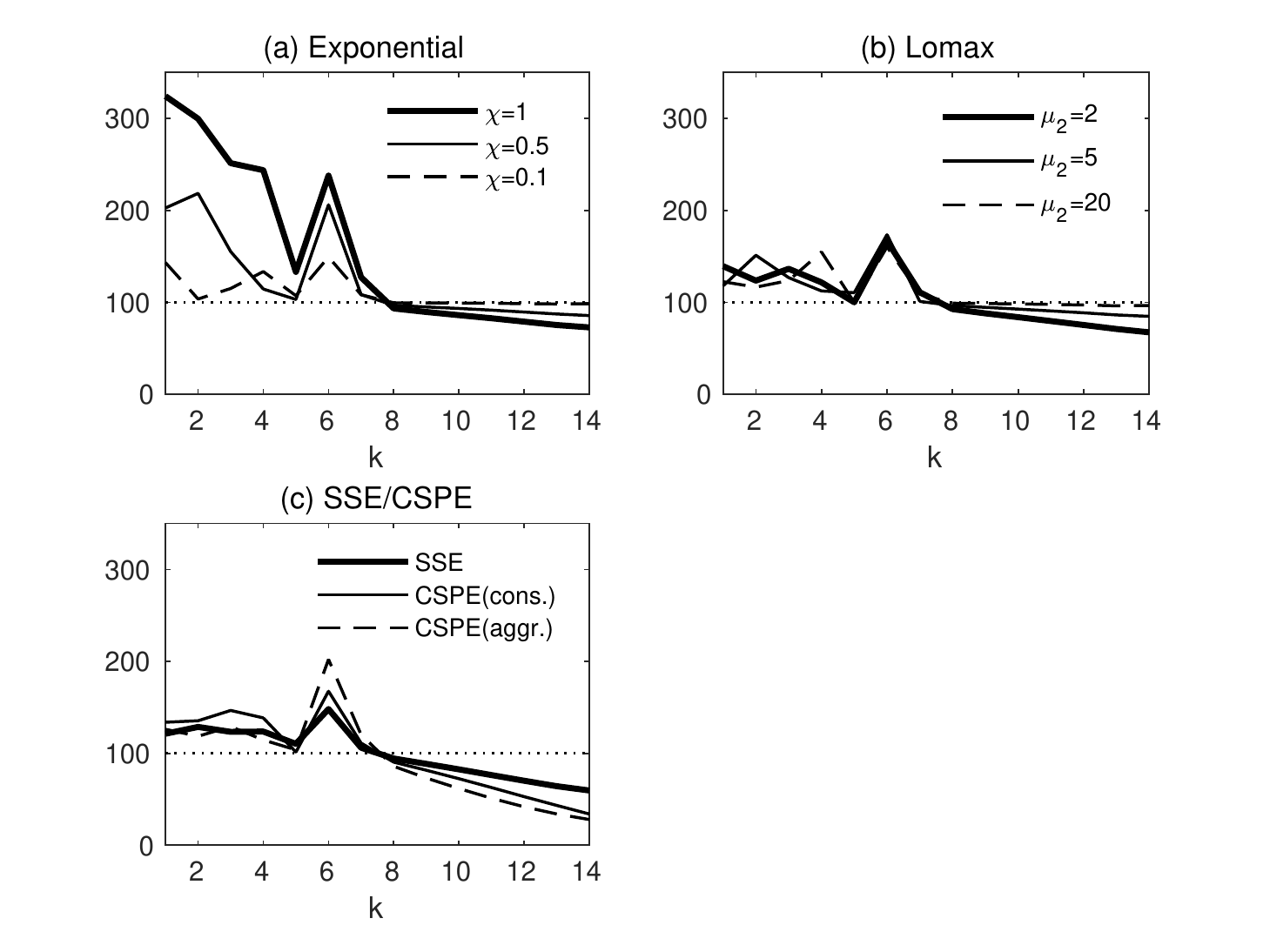} 
\end{figure}

\clearpage{}

\begin{figure}
\caption{Aggregate industrial production growth}

\centering{}\includegraphics{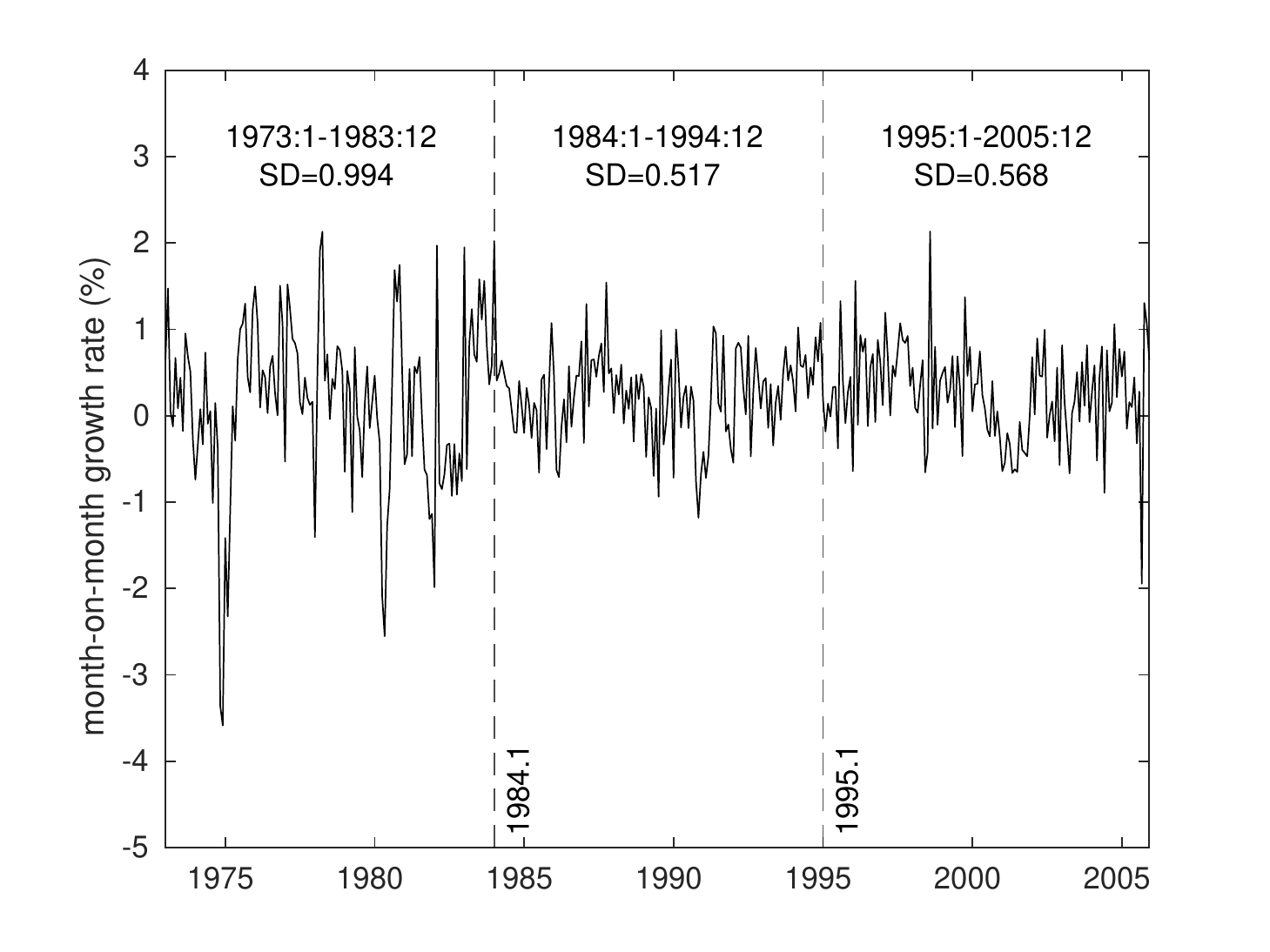} 
\end{figure}

\clearpage{}

\begin{figure}
\caption{Volatility of sectoral industrial production growth}

\centering{}\includegraphics[scale=1.2]{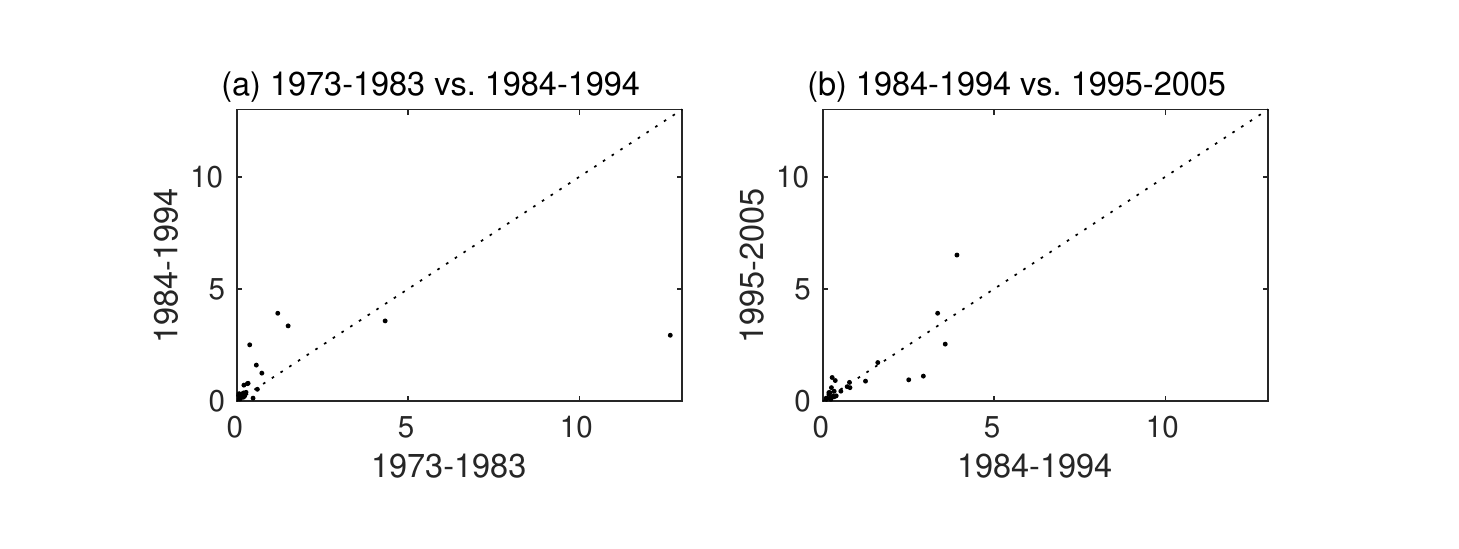} 
\end{figure}

\clearpage{}

\begin{figure}
\caption{Posterior estimates of $\boldsymbol{\omega}$}

\begin{centering}
\includegraphics[scale=1.2]{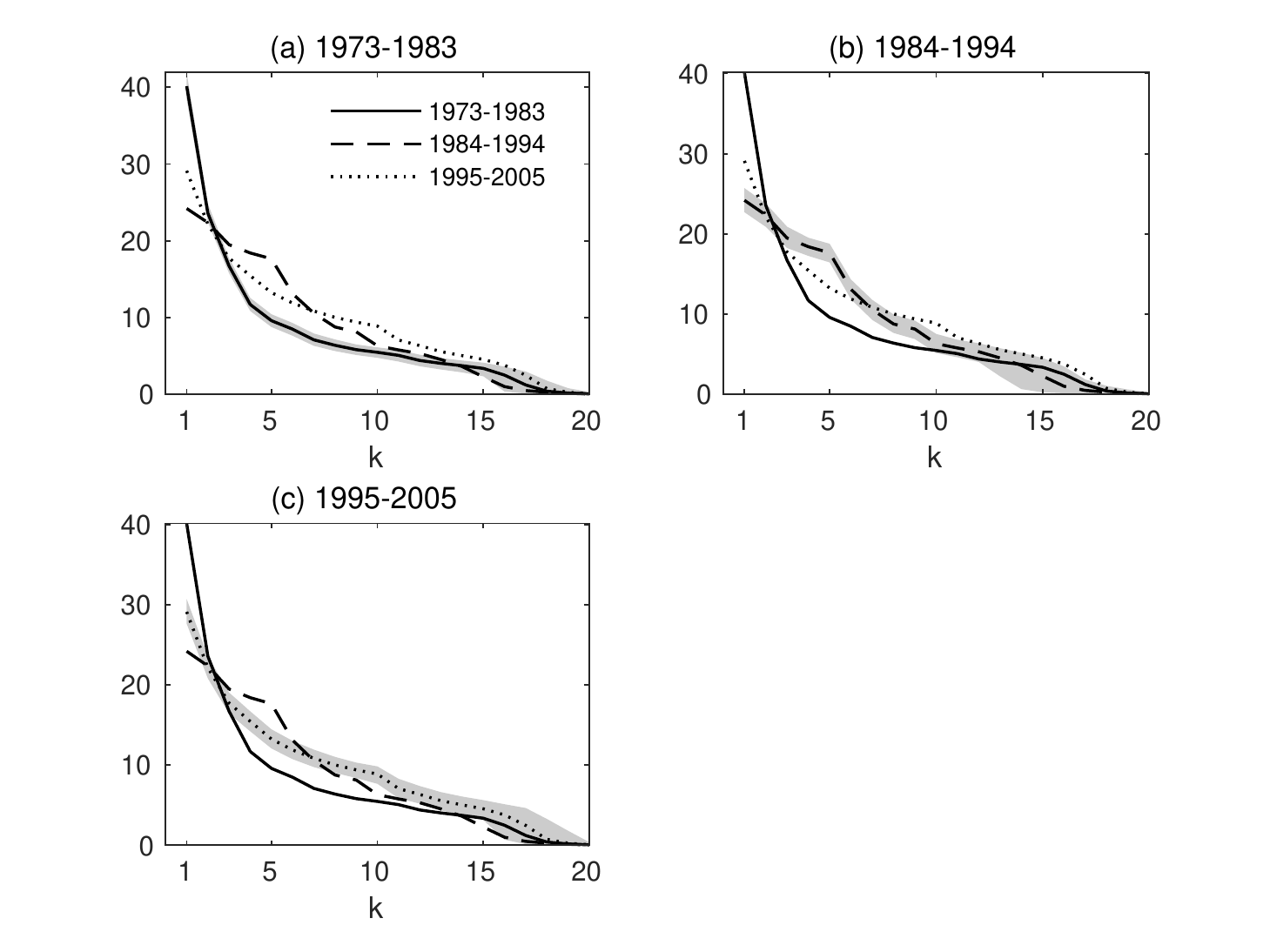} 
\par\end{centering}
\centering{}%
\begin{minipage}[t]{0.8\columnwidth}%
Note: Solid, dashed, and dotted lines indicate posterior mean estimates.
The same lines are plotted in the three panels. The shaded areas indicate
the 95\% credible sets for each period.%
\end{minipage}
\end{figure}

\clearpage{}

\begin{figure}
\caption{Posterior density of $\tau^{-1}$}

\centering{}\includegraphics{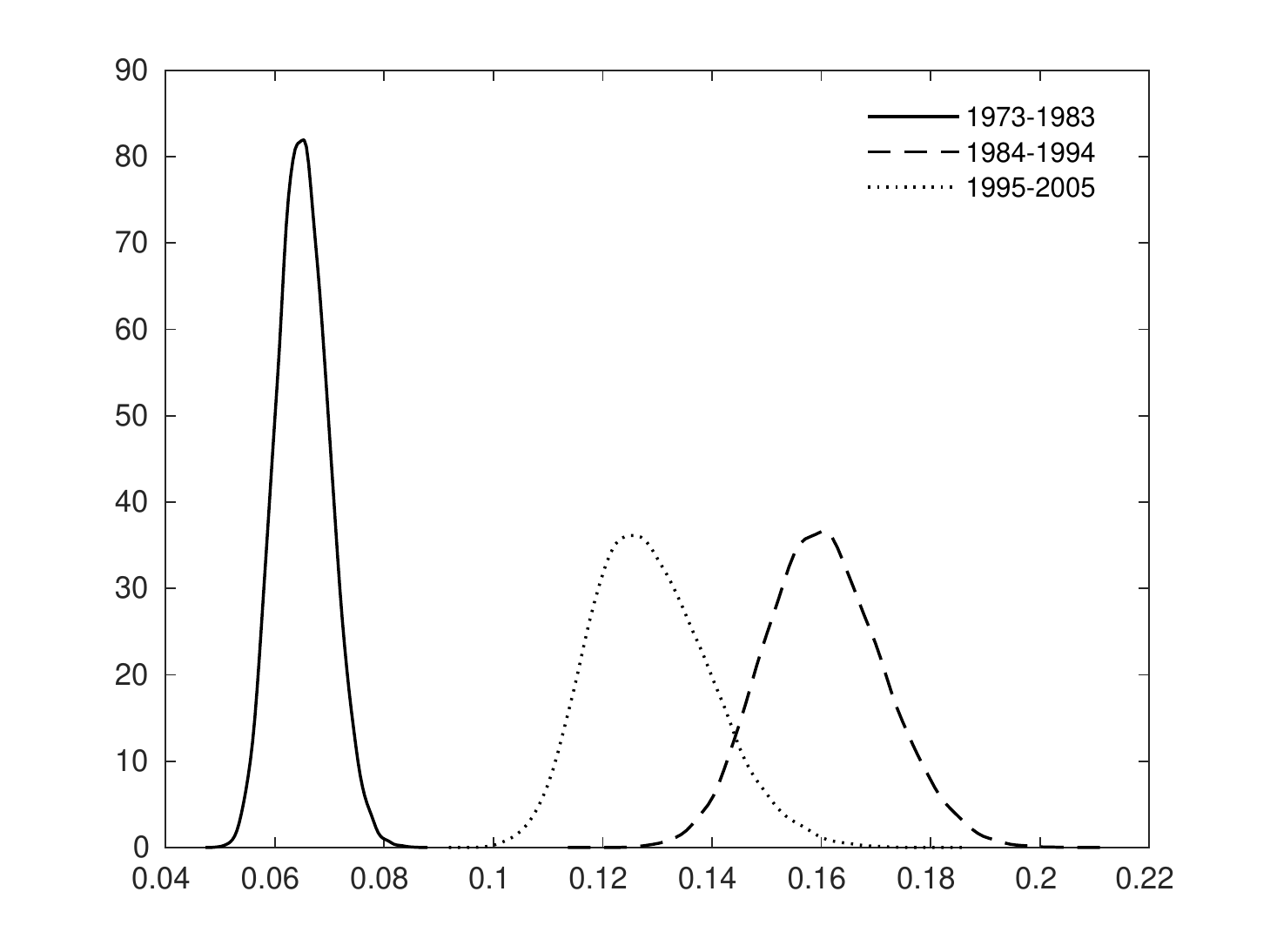} 
\end{figure}

\clearpage{}

\begin{figure}
\caption{Change in variance contribution}

\begin{centering}
\includegraphics[scale=1.2]{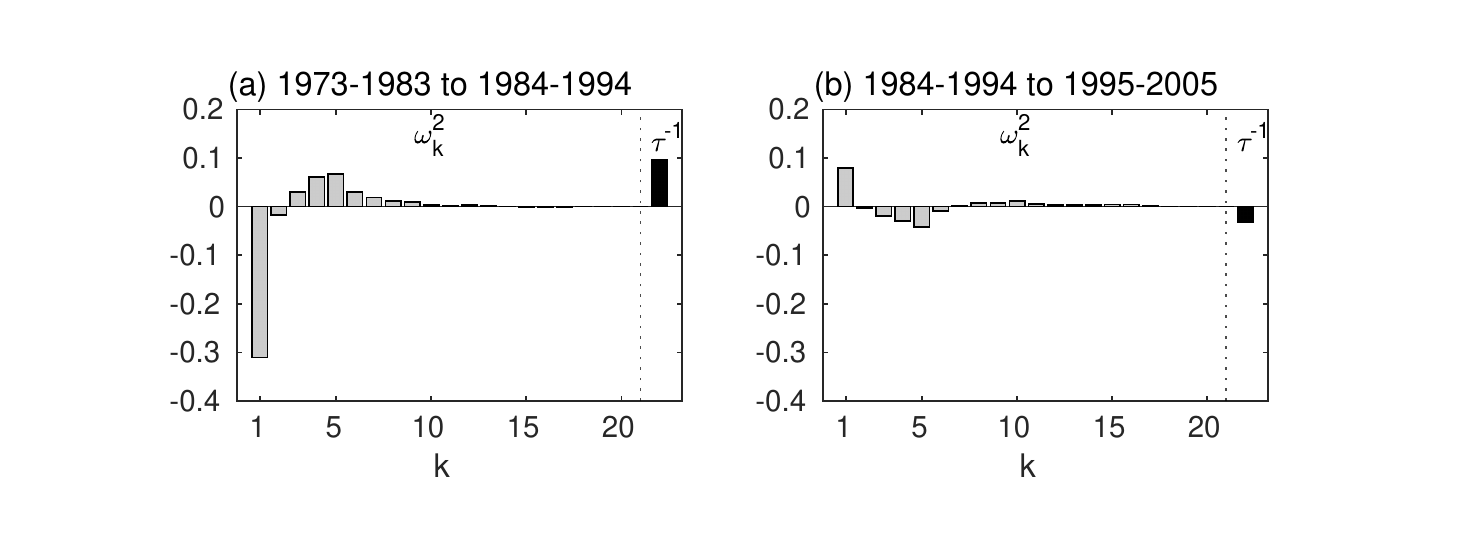} 
\par\end{centering}
\centering{}%
\begin{minipage}[t]{0.8\columnwidth}%
Note: Shaded and black bars, respectively, indicate the differences
in the posterior mean estimates of $\omega_{k}^{2}$ and $\tau^{-1}$
for different periods.%
\end{minipage}
\end{figure}

\end{document}